\documentclass[aps,pra,reprint,superscriptaddress]{revtex4-2}
\usepackage{graphicx}
\usepackage{dcolumn}
\usepackage{bm}
\usepackage{subfig}  
\usepackage{caption}  
\usepackage{ragged2e} 
\usepackage{hyperref}
\usepackage{amsmath}
\usepackage{esint}
\usepackage{float}
\usepackage{orcidlink}

\begin{document}
	
\preprint{APS/123-QED}
	
\title{CR-39 track detector signatures of slow neutron like signals in Heavy-water electrolysis}
	
\author{Ankit Kumar\,\orcidlink{0000-0001-5962-7914}}
\email{ankitkumar5111994@gmail.com, ankt@iitk.ac.in}
\affiliation{Department of Physics, Indian Institute of Technology Kanpur, Kanpur, India}
	
\author{Tushar Verma\,\orcidlink{0009-0005-1810-6561}}
\affiliation{Department of Chemical Engineering, Indian Institute of Technology Kanpur, Kanpur, India}
	
\author{Pankaj Jain\,\orcidlink{0000-0001-8181-5639}}
\affiliation{Space, Planetary \& Astronomical Sciences \& Engineering, Indian Institute of Technology Kanpur, Kanpur, India}
	
\author{Raj Ganesh Pala\,\orcidlink{0000-0001-5243-487X}}
\affiliation{Department of Chemical Engineering, Indian Institute of Technology Kanpur, Kanpur, India}
	
\author{K.~P.~Rajeev\,\orcidlink{0000-0002-4685-6766}}
\email{kpraj@iitk.ac.in}
\affiliation{Department of Physics, Indian Institute of Technology Kanpur, Kanpur, India}

\begin{abstract}
We report reproducible track-detector signals  consistent with slow neutron capture events, recorded in D$_2$O electrolysis involving D-Pd deposited on Pt cathode. Sensitivity to slow neutrons was achieved using boron-coated CR-39 (BCR) detectors, which register charged particle tracks arising from the $^{10}$B$(n,\alpha)^{7}$Li reaction. These detectors were positioned adjacent to identically prepared uncoated CR-39 control detectors (CCR), which are effectively insensitive to slow neutrons and serve to quantify background contributions from charged particles and fast neutrons under the present experimental conditions. A reproducible differential detector signature (BCR $>$ CCR) would thus indicative of slow neutron fluences.  
Across multiple independent D$_2$O electrolysis experiments in $0.25~\mathrm{T}$ field, the BCR exhibited significantly excess track signals relative to CCRs. Under these conditions, the observed differential response corresponds to an inferred detector-equivalent slow neutron flux of approximately $(6.7 \pm 0.2)~\mathrm{cm^{-2},s^{-1}}$. Removal of the magnetic field resulted in a reduction of the differential signal by a factor of $\sim6$, indicating a strong empirical dependence on the applied field. In contrast, H$_2$O electrolysis performed under otherwise identical conditions produced no measurable differential detector response, establishing the necessity of deuterated electrochemical conditions for the observed effect.
The results are reported strictly as detector signatures consistent with slow neutron capture and do not assert any theoretical explanation. Instead, this work establishes a control verified and detector validated experimental protocol for detecting low flux slow neutrons, and provides empirical constraints relevant to slow neutron studies in experiments involving metal-deuteride systems.
\end{abstract}

\maketitle
\section{Introduction}

Understanding how condensed-matter environments can influence nuclear reaction rates under low-energy conditions remains an open question of importance in nuclear physics. Electrochemical experiments on metal deuterides have reported the detection of charged-particle emission signatures using solid-state nuclear track detectors~\cite{gotzmer2023li,mosier2017detection,roussetski2017detection,mosier2009characterization,mosier2007use,szpak2005evidence,oriani2002generation}. During heavy-water electrolysis, deuterons are co-deposited with palladium onto metal cathodes, forming metal deuteride systems~\cite{szpak1995cyclic,szpak1994deuterium,fukai2005metal} that exhibit such emissions, with enhanced rates observed under externally applied magnetic or electric fields~\cite{mosier2007use,US8419919B1}.

Neutron production has also been explored using He-3 detectors in such experiments, but the detected signals were not consistently above the background~\cite{gotzmer2023li}. Thus, definitive assessment of neutron emission was suppressed by the limitations of active neutron detectors (He-3 in this case), including background instabilities, the need for neutron moderation, and susceptibility to electromagnetic interference in electrochemical cells, which collectively reduce its sensitivity at low flux.

The present work seeks detector signatures consistent with slow-neutron capture during electrochemically prepared metal deuterides by employing a CR-39 (Columbia Resin 39) based differential detection method that separates slow-neutron interactions from charged-particle and fast-neutron backgrounds.

CR-39 is a solid-state nuclear track detector~\cite{durrani2013solid} that records incident charged nuclear particle that deposits at least $\approx15$~keV~$\mu$m$^{-1}$ (linear energy transfer (LET) threshold for Tastrak CR-39)~\cite{jadrnickova2006dosimetry}, with near-unity efficiency above the LET threshold. Its detection principle is based on fundamental ion-matter interactions, in which an incident charged particle traversing the CR-39 polymer produces a permanent latent damage trail along its path, which is subsequently revealed as a well-defined micro-pit (track opening) on the CR-39 surface after chemical etching, and marks the positions of charged-particle interactions. Additionally, the track parameters can be used to characterize the charge and mass of the incident ion~\cite{durrani2013solid,schollmeier2023differentiating}. When CR-39 is exposed to a charged-particle field, the micro-pit areal track density serves as a deterministic representative (quantitative proxy) of the incident flux at the detection site~\cite{durrani2013solid,fleischer2022nuclear}. CR-39 is thus a passive detector that eliminates electronics-based background instabilities, provides immunity against electric and/or magnetic fields, and is robust enough to be utilized in complex environments such as liquid media or plasma experiments. Of particular relevance to the present work, CR-39 exhibits negligible sensitivity to slow neutrons, gamma photons, and beta particles, since these radiations deposit energy below the linear energy transfer threshold required to produce latent damage trails in the polymer~\cite{durrani2013solid,fleischer2022nuclear}.

To investigate potential slow neutrons signatures in electrolysis experiments, we employ boric-acid-coated CR-39 (BCR) detectors. BCR records tracks produced by the charged reaction products ($\alpha$ particles and $^{7}$Li nuclei) generated via the well-characterized $^{10}$B$(n,\alpha)^{7}$Li capture reaction~\cite{knoll2010radiation}, which is sensitive to neutrons in the thermal and epithermal energy range (referred to as slow neutrons in this paper). To confirm that the BCR response originates only from slow neutrons, a CR-39 chip without boron coating (CCR) was positioned adjacent to the BCR. This control detector does not respond to slow neutrons and therefore should record tracks arising from all non-slow-neutron background contributions, including charged particles and fast neutrons, if present. Accordingly, this methodology does not provide independent neutron energy spectroscopy.

Performance wise, CR-39 has been demonstrated to be an effective and passive neutron dosimeter across a broad energy range, from thermal ($\sim25$~meV) to fast ($\sim66$~MeV) neutrons~\cite{izerrouken2003wide}, and has been widely used in applications such as thermal-neutron dosimetry~\cite{nassiri2025optimization}, boron neutron capture studies~\cite{smilgys2013boron}, and D--D fusion experiments~\cite{PhysRevLett.112.095001}. Of particular relevance to the present work, CR-39 was also employed in a study reporting thermal neutrons in high-voltage atmospheric discharge events via the $^{10}$B$(n,\alpha)^{7}$Li reaction~\cite{agafonov2013observation}.

The characteristics and performance of the CR-39 detectors were independently verified in advance and are documented in the Supplemental Material. These validations demonstrate the suitability of the detector for slow-neutron detection and flux measurements, particularly in the low-flux regime relevant to the present study. In the following sections, we present a systematic investigation employing this diagnostic during D$_2$O electrolysis, including multiple independent repetitions, extensive background measurements, a magnet off control, and a light-water (H$_2$O) negative control to exclude alternative interpretations of the experimental results. Strict chain-of-custody procedures for detector handling were implemented, since CR-39, like other passive detectors, cannot be reset once exposed, and are provided in the Supplemental Material.

The goal of this work is not to propose a theoretical mechanism but to present a reproducible, detector verified measurement protocol capable of determining whether slow neutron like signatures can be observed under the tested conditions. The results provide an empirical basis for further inquiry into the nuclear physics involved in slow neutron like signals generation in deuterated metals and establish a reference experimental platform for future nuclear diagnostic studies in low flux environments.

\section{Experimental Setup}
\begin{figure}[!tb]
	\centering
	\includegraphics[width=\columnwidth]{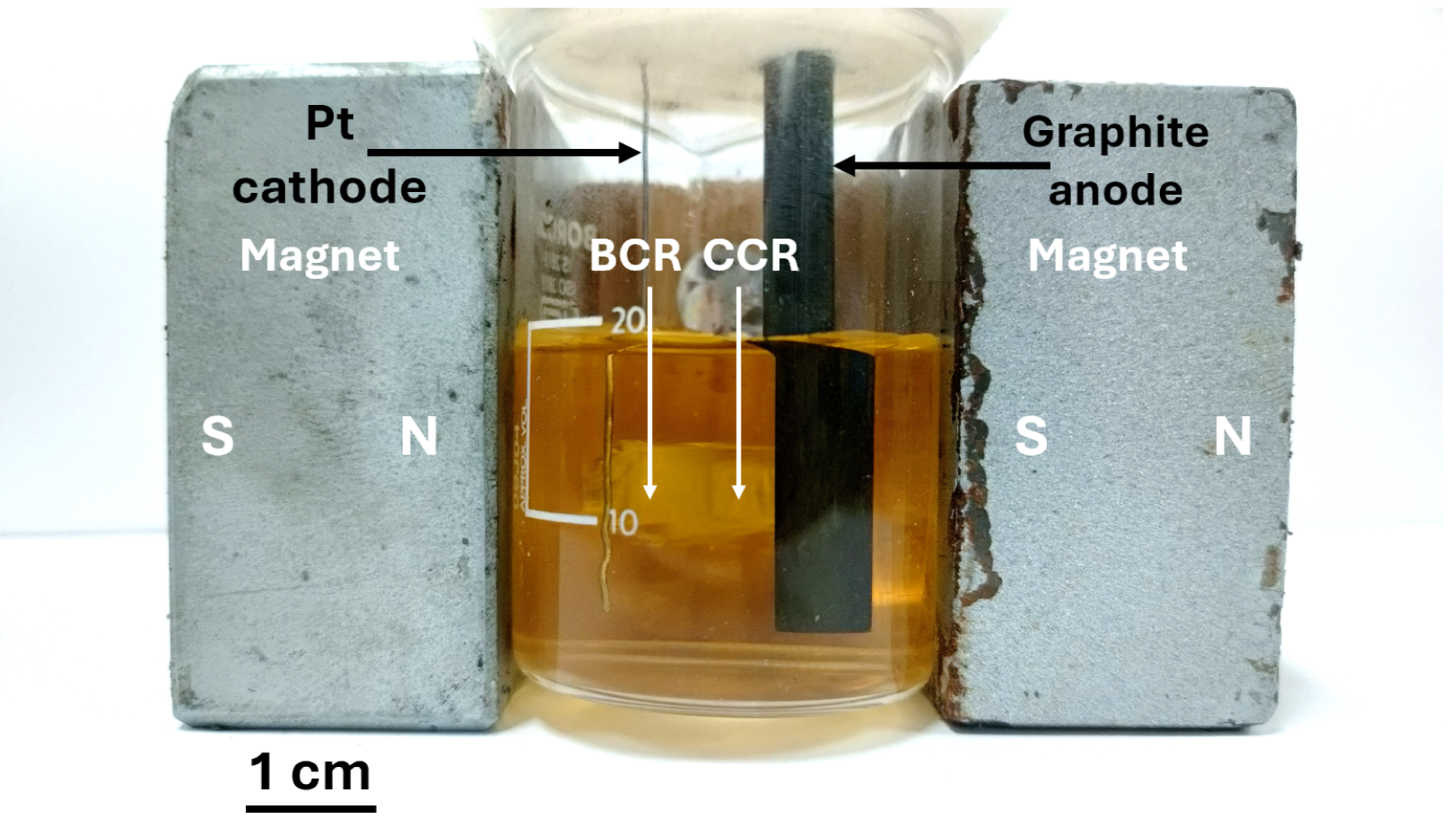}
	\caption{\justifying Electrolytic cell placed in external magnetic field.}
	\label{fig:1}
\end{figure}

The experiment uses an electrolytic cell ($25$~mL beaker) positioned between two neodymium magnets that generate a $0.25 \pm 0.01$~T magnetic field directed from the Pt cathode to the graphite anode, with the cell containing $20$~mL of an electrolyte composed of $1.25$~mM PdCl$_2$ and $150$~mM LiCl in D$_2$O [Fig.~\ref{fig:1}]. 
A boron-coated and an uncoated CR-39 (BCR and CCR), both encapsulated in $65~\mu$m thick plastic polythene, were immersed in the electrolyte and fixed inside the beaker, facing the electrodes [see Fig.~S1 in Supplemental Material].


Electrolysis was conducted at $50.0 \pm 0.1~\mu$A for 3 days, followed by $100 \pm 5$~mA ($7.0 \pm 0.5$~V) for 4 days. 
To compensate for electrolyte loss due to gas evolution, pure D$_2$O was replenished, as required, daily. 
After electrolysis, the CR-39 substrates of BCR and CCR were retrieved and etched in $6$~M NaOH at $70 \pm 2$~$^\circ$C for 6 hours to reveal the particle tracks recorded on them during electrolysis. 
From a microscopic analysis of the etched CR-39, a mean track count on each detector type was obtained, and two corrections were applied to obtain the track counts recorded during the electrolysis run: (1) subtraction of baseline tracks, defined as the tracks present on CR-39 prior to any exposure; and (2) subtraction of background tracks, defined as the tracks observed on BCR and CCR in the electrolytic cell under electrolysis-OFF conditions while the magnetic field was applied. 
The resulting BCR and CCR track counts were then compared and analyzed. 
Control experiments were also performed, including heavy-water (D$_2$O) electrolysis in the absence of an applied magnetic field and light-water (H$_2$O) electrolysis in the presence of a magnetic field, to support the conclusions of this study. 
The details of the experimental methods are provided in Sec.~I of the Supplemental Material.

\section{Results and Discussions}
Microscopic analysis of detectors utilized in D$_2$O electrolysis experiments under a $0.25$~T field revealed significantly higher particle tracks on BCR compared to the adjacent CCR.


\begin{figure*}[!tb]
	\centering
	\includegraphics[width=\textwidth]{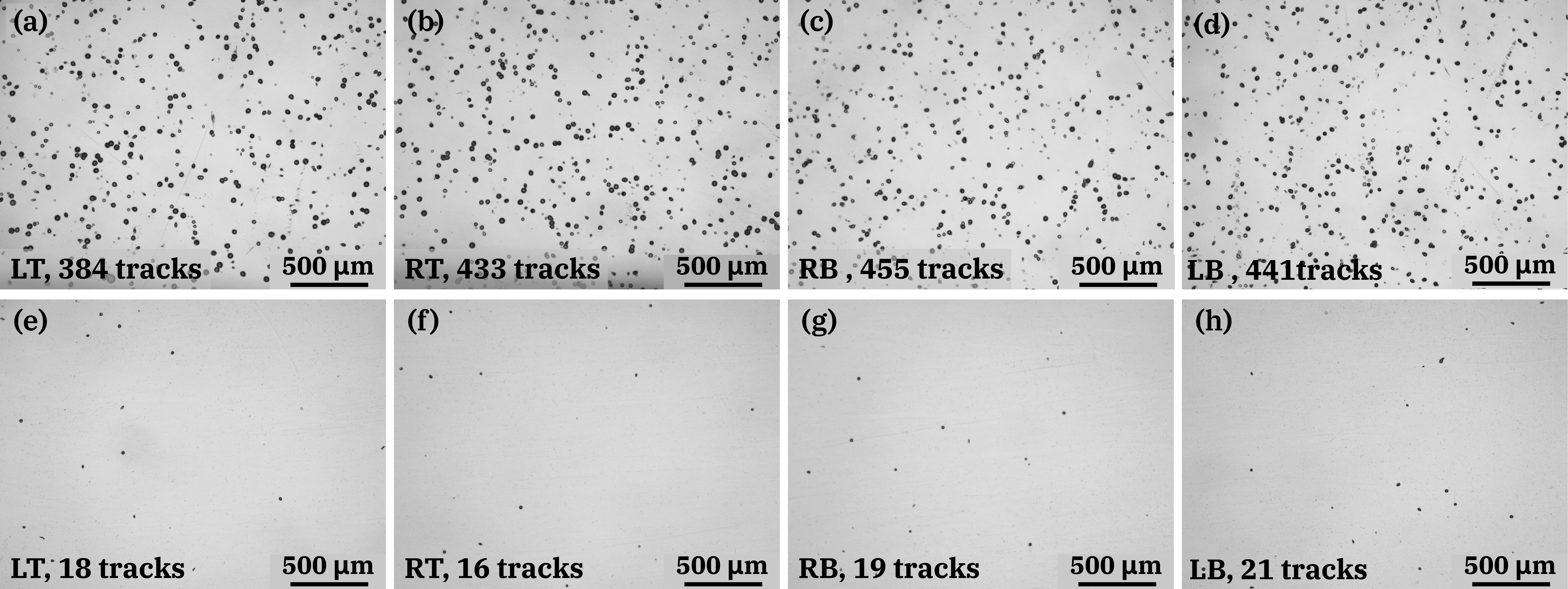}
	\caption{\justifying Optical micrographs of etched tracks on boron-coated CR-39 (BCR) and control CR-39 (CCR) detectors exposed during D$_2$O electrolysis under a 0.25~T magnetic field. For each detector (6~mm $\times$ 4~mm), four equal non overlapping regions, each of area 1.8~mm $\times$ 2.5~mm (4.8~mm$^2$) were imaged: left-top (LT), right-top (RT), right-bottom (RB), and left-bottom (LB). Panels (a)–(d) show BCR images with track counts of 384 (LT), 433 (RT), 455 (RB), and 441 (LB), yielding a mean of $428 \pm 13$ tracks per 4.8~mm$^2$; after subtraction of the baseline ($11.6 \pm 0.7$ tracks) and BCR background ($3.39 \pm 0.10$ tracks), the net BCR signal recorded during electrolysis is $413 \pm 13$ tracks per 4.8~mm$^2$. Panels (e)–(h) show the corresponding CCR images with track counts of 18 (LT), 16 (RT), 19 (RB), and 21 (LB), yielding a mean of $18.5 \pm 0.9$ tracks per 4.8~mm$^2$; after subtraction of the baseline and CCR background ($1.25 \pm 0.11$ tracks), the net signal recorded by the CCR during electrolysis is $5.7 \pm 1.2$ tracks per 4.8~mm$^2$. Raw data for baseline and background determinations are provided in Secs.~II~A–C of the Supplemental Material.}
	\label{fig:2}
\end{figure*}

As shown in Figs.~\ref{fig:2}(a-d), four surveyed regions on BCR, after baseline and background correction, recorded an average of $413 \pm 13$ tracks ($\sim 121\times$ the BCR background). Similarly, the adjacently placed CCR [see Figs.~\ref{fig:2}(e-h)] showed an average of $5.7 \pm 1.2$ tracks. Thus, both detectors recorded measurable activity, with the BCR exhibiting an excess of $\sim 407 \pm 13$ tracks, larger by a factor of $\sim 71$ relative to the tracks recorded by CCR.

In general, a CCR has higher detection efficiency than the BCR for any incident charged particle (having enough kinetic energy to cross the $65~\mu$m thick polythene layer and create a track in the underlying CR-39 substrate) or fast neutrons, since the boron layer of the BCR scatters these particles while the CCR allows a direct interaction with the CR-39. Accordingly, the BCR should always register fewer tracks than the CCR if no nuclear reaction is involved with the boron layer. The tracks observed on the CCR ($5.7 \pm 1.2$ tracks) can therefore be attributed to these mechanisms, with the adjacently placed BCR expected to record at most the same or fewer. We argue that the only case where the BCR will record more tracks than the CCR is exposure to a slow-neutron field. Capture of slow neutrons by $^{10}$B produces $\alpha$ and $^{7}$Li particles via $^{10}$B(n, $\alpha$)$^{7}$Li, which form tracks in the underlying CR-39. As the CCR contains no boron, no such response is possible, making slow neutron capture the unique explanation for the observed BCR excess of $407 \pm 13$ tracks over the CCR, fully consistent with neutron--boron interactions in the coating. Thus, the CCR sets an upper limit on all non-slow-neutron contributions by putting a bound on the possible fast-neutron and charged-particle backgrounds and further validates that the large BCR excess originates from neutron--boron reactions during electrolysis.

Next, to evaluate the variation of particle track signals on BCRs and CCRs during D$_2$O electrolysis in presence of magnetic field, the experiment was independently repeated four more times besides the one discussed above, with new electrodes and freshly prepared chemical solutions used for each run.


\begin{figure}[!thbp]
	\centering
	\includegraphics[width=\columnwidth]{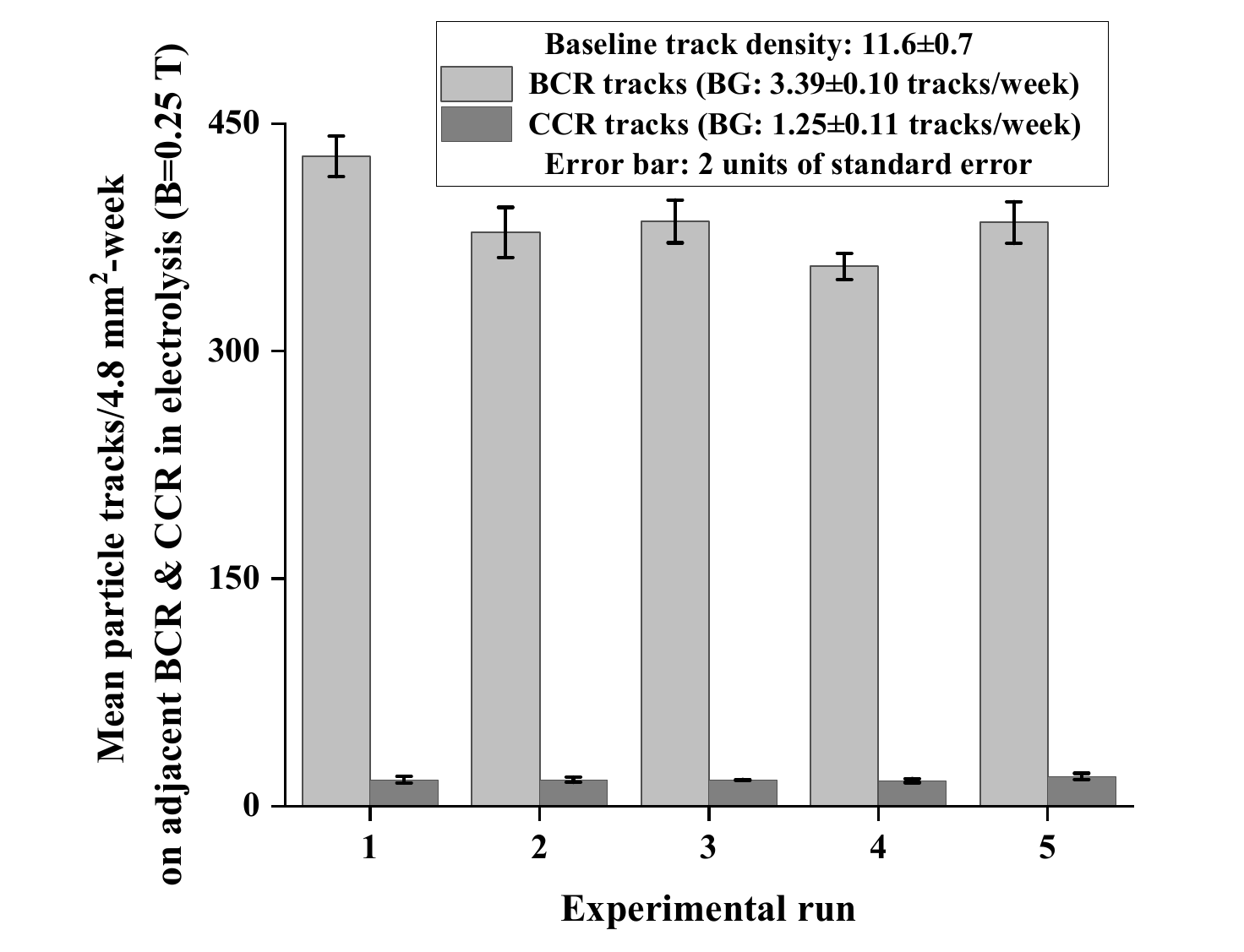}
	\caption{\justifying Mean particle-track densities measured on boron-coated CR-39 (BCR) and control CR-39 (CCR) detectors for five independent experimental runs conducted during D$_2$O electrolysis under a 0.25~T magnetic field. For each run, the plotted values represent the mean track count per 4.8~mm$^2$ obtained from the four spatial regions of a given detector (LT, RT, RB, and LB), with paired BCR and CCR measurements acquired under identical conditions; no baseline or background subtraction was applied. Bar~1 corresponds to the dataset discussed previously, while bars~2--5 represent additional repeated trials performed under identical conditions. Further details of the bar-graph construction are provided in Sec.~II~D of the Supplemental Material.}
	\label{fig:3}
\end{figure}

Figure~\ref{fig:3} presents the mean particle-tracks measured on BCR and CCR detectors for five independent D$_2$O electrolysis under  0.25~T in the form of bar graph (bar 1 shows the previously discussed result). Across all five experimental runs, the BCR consistently recorded track densities exceeding those of the adjacent CCR detectors. After baseline and background subtraction, the BCR track counts ranged from $341 \pm 9$ to $413 \pm 13$ tracks, with an average BCR track counts of $372 \pm 11$ tracks per week. In contrast, the corresponding mean CCR track counts remained consistently low, typically below 7 tracks per week. A Welch two-sample $t$-test performed over the BCR and CCR values observed in five trials yielded a $p$-value $\ll 0.001$, indicating a statistically significant difference between the two datasets. [Details of the calculation are provided in Sec.~II~E of the Supplemental Material]. These results confirm that the experimental setup is well controlled, the procedures are consistent, and the observed effects are not random or single-run anomalies. 

To isolate the effect of the magnetic field on particle detector signals, an additional D$_2$O electrolysis experiment was performed without magnets, while keeping all other electrochemical conditions identical.


\begin{figure*}[!htbp]
	\centering
	\includegraphics[width=\textwidth]{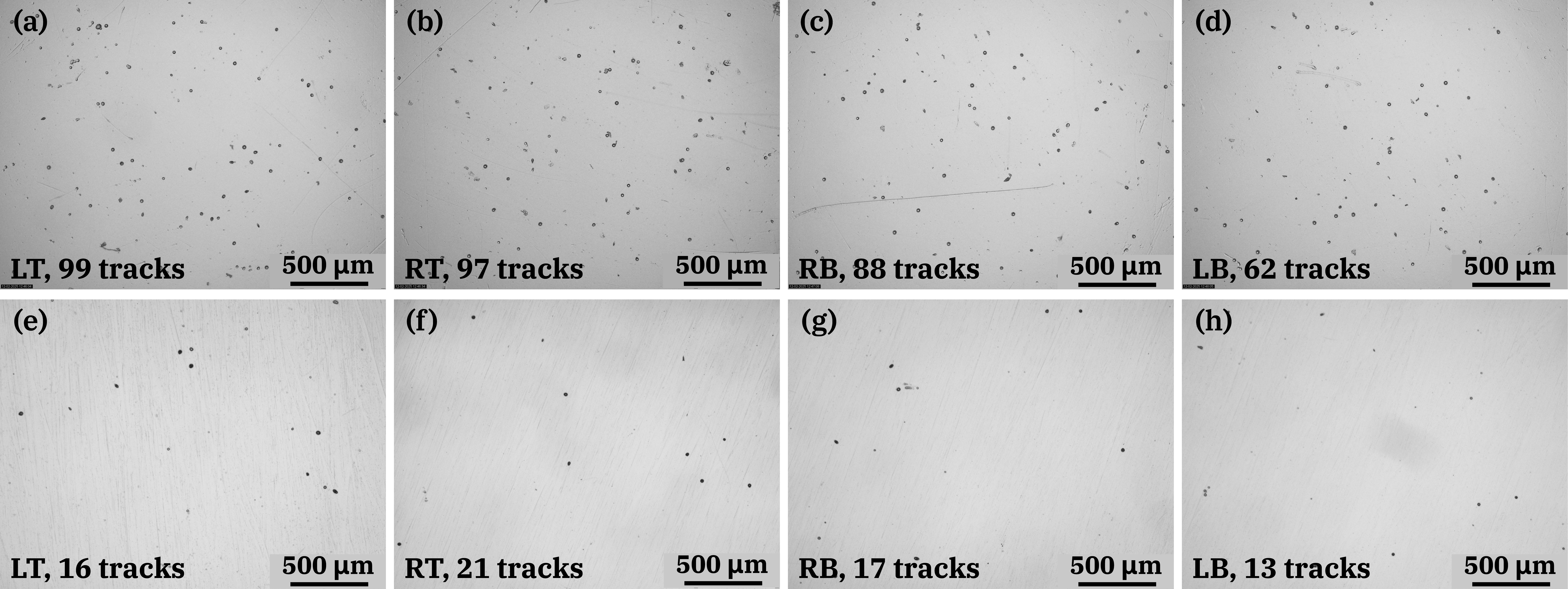}
	\caption{\justifying Microscopic images showing particle tracks recorded on detectors in D$_2$O electrolysis conducted without a magnetic field. BCR detector shows track counts on: (a) LT-99, (b) RT-97, (c) RB-88 and (d) LB-62 tracks, yielding an average of $87 \pm 7$ tracks per 4.8~mm$^2$, which after subtracting the baseline and BCR background, amounts to $72 \pm 7$ tracks per 4.8~mm$^2$. CCR detector regions show track counts on (e) LT-16, (f) RT-21, (g) RB-17 and (h) LB-13 regions, yielding a mean of $16.8 \pm 1.4$ tracks per 4.8~mm$^2$, which after baseline and weekly background corrections, corresponds to a residual CCR signal of $4.0 \pm 1.6$ tracks per 4.8~mm$^2$. [For raw images, see Sec.~II~F in Supplemental Material].}
	\label{fig:4}
\end{figure*}

\begin{figure*}[!htbp]
	\centering
	\includegraphics[width=\textwidth]{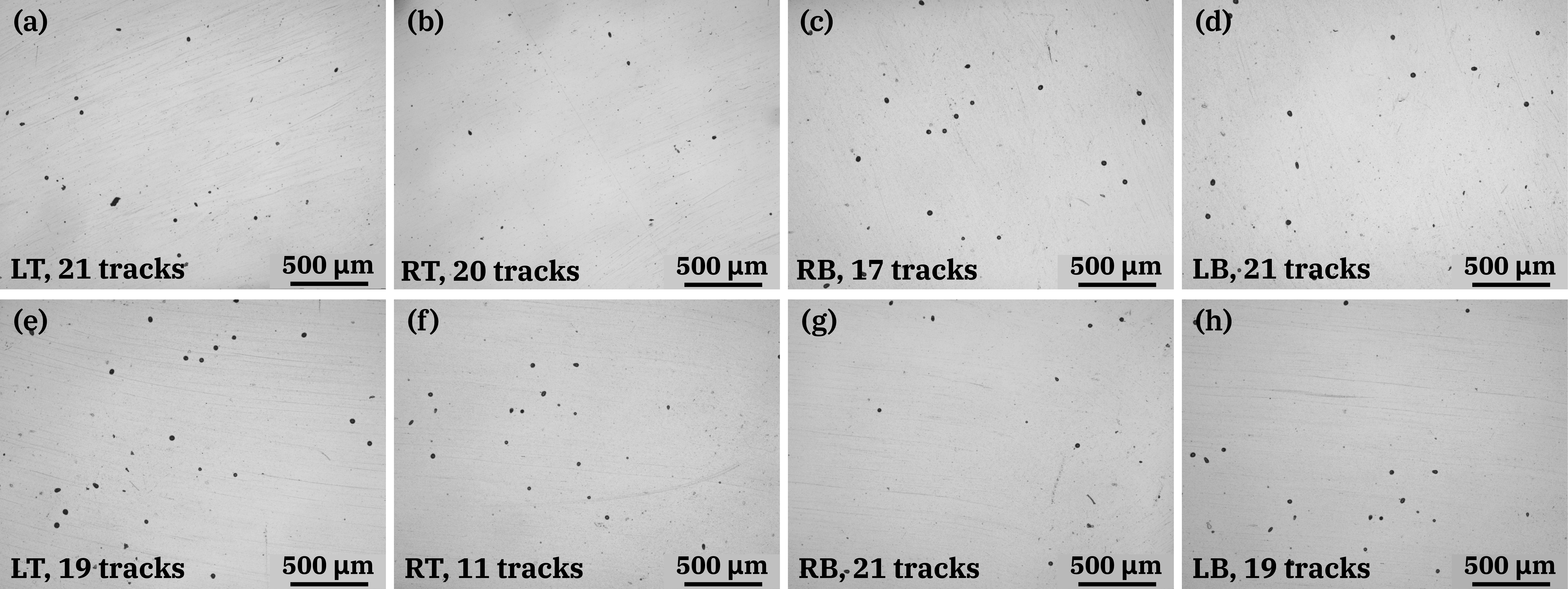}
	\caption{\justifying Microscopic images showing particle tracks recorded on detectors in H$_2$O electrolysis conducted in a magnetic field. BCR detector shows track counts on: (a) LT-21 tracks, (b) RT-20, (c) RB-17 and (d) LB-21 regions, yielding an average of $19.8 \pm 0.8$ per 4.8~mm$^2$ tracks, which after subtracting the baseline and BCR background, amounts to $4.8 \pm 1.1$ tracks per 4.8~mm$^2$. CCR detector regions show track counts on (e) LT-19, (f) RT-11, (g) RB-21 and (h) LB-19 regions, yielding a mean of $17.5 \pm 1.9$ tracks per 4.8~mm$^2$, which after baseline and weekly background corrections, corresponds to a residual signal of $4.7 \pm 2.0$ tracks per 4.8~mm$^2$ [For raw images, see Sec.~II~K in the Supplemental Material].}
	\label{fig:5}
\end{figure*}

In the absence of a magnetic field, the BCR recorded $72 \pm 7$ tracks [Figs.~\ref{fig:4}(a-d)] and the CCR recorded $4.0 \pm 1.6$ tracks [Figs.~\ref{fig:4}(e–h)]. The data shows that in the absence of magnetic field the BCR recorded an excess of 68 tracks over CCR, indicating significant number of slow neutron signals, despite being reduced by a factor of $\sim6$ compared to the mean BCR tracks obtained in presence of magnetic field as presented in Fig.~\ref{fig:3}. A two-sample Welch $t$-test on the BCR and CCR dataset obtained in this case gives $0.001 < p < 0.002$, which indicates a statistically significant difference between the BCR and CCR values [for calculation details, see Sec.~II~G in Supplemental Material]. We note that the ambient laboratory background accumulated on bare CR-39 during detector handling ($\le 5$ min), contributes only ($0.047 \pm 0.002$) tracks per $4.8$~mm$^2$ over 5 min and is therefore a negligible factor in track analysis of electrolysis experiments (see Sec.~II~H of the Supplemental Material for details). It is also important to note that the detection methodology used here does not discriminate between primary slow neutrons and fast neutrons that may have been thermalized through moderation in the heavy-water medium prior to detection by the BCR. Furthermore, we emphasize that the observed dependence at a single magnetic-field value reported here is purely empirical. No causal relationship between the applied magnetic field and nuclear interactions has been established, and the underlying mechanism remains unresolved. Additional experiments involving systematic variation of the magnetic-field strength, aimed at identifying any possible scaling behavior of the neutron emission, are currently underway.

The detection efficiency (probability), $\varepsilon$ of BCR was measured to be $(1.91 \pm 0.02) \times 10^{-3}$ using a thermal neutron reference field~\cite{szpak2005evidence}. Using this detection efficiency, the slow neutron flux in D$_2$O electrolysis in 0.25 T field, at BCR detection site was determined to be $\Phi$ = $(6.7 \pm 0.2)\, \mathrm{cm^{-2}\,s^{-1}}$. Details of the calibration procedure and flux calculations are provided in Sec.~II~I and K of the Supplemental Material. 

An order-of-magnitude estimate can be made of the activity of a hypothetical linear radioactive wire source placed at the center of the cell that would generate a neutron flux at the BCR detector comparable to that observed during D$_2$O electrolysis. Under simplifying assumptions, as detailed in Sec.~II~J of the Supplemental Material, this detector-equivalent activity is estimated to be $(6.8 \pm 0.2)$~nCi. This value does not represent a physical source strength but serves solely as a detector-equivalent normalization for internal comparison.

As a negative control experiment, an identical set of electrolysis experiment is performed in light water (H$_2$O) in place of D$_2$O wherein metal hydride is formed over Pt cathode.


In the H$_2$O control set as shown in Figure~\ref{fig:5}, BCR recorded $4.8 \pm 1.1$ tracks, while the adjacent CCR remained at $4.7 \pm 2.0$ tracks, more or less indistinguishable from each other and more than two orders of magnitude lower than the mean D$_2$O response ($372 \pm 11$ tracks) in presence of 0.25~T, as presented in Fig.~\ref{fig:3}. The absence of any measurable BCR excess in H$_2$O electrolysis strongly supports the conclusion that the neutron-induced tracks observed in D$_2$O arise from a deuterium-dependent process, rather than from detector artifacts or environmental backgrounds.

Section~II~L of the Supplemental Material shows that plastic shielding results in complete attenuation of $\alpha$ particles with energies up to 5.48~MeV on CR-39 detectors. The insensitivity of CR-39 to $\beta$ and $\gamma$ radiation is also demonstrated therein.

\subsection{Discussion of Mechanisms and its Implications}
The present work does not establish any specific microscopic mechanism, and no accepted theoretical framework currently explains nuclear reactions at such low energies. Some studies propose that second-order processes involving resonant intermediate states may enhance low-energy nuclear transitions~\cite{kumar2025resonant,ramkumar2024low}, with energy exchange between a D--D pair and the lattice via Coulomb scattering~\cite{kalman2019forbidden,jain2024medium} or photon emission~\cite{ramkumar2024low,kumar2025resonant,jain2022photon,jain2022low}. If a low-energy D--D resonance exists and is accessible under electrochemical conditions~\cite{dubey2025experimental,czerski2022deuteron}, such mechanisms could enhance neutron-producing channels relative to direct transitions~\cite{kumar2025resonant}. These scenarios remain speculative. Regarding the influence of magnetic fields on the enhanced detector response observed during D$_2$O electrolysis, magnetic fields are known to affect hydrogen and deuterium evolution kinetics at cathode through established electrochemical mechanisms~\cite{vensaus2024enhancement,belgami2024exploring}. However, the observed 6 times increase in BCR signals at $0.25$~T over $0$~T cannot be readily explained by these effects alone.

\subsection{Experimental implications and summary}
Our results advances previously reported D$_2$O electrolysis studies reporting enhanced charged particle emissions from Pd/D co-deposited cathodes under an externally applied magnetic field. ~\cite{smith2021electrolytic,mosier2009triple,szpak2005evidence,szpak2007further,mosier2007use,mosier2008reply}. In alignment with these observations, the present work demonstrates that a measurable slow-neutron induced particle track signals are generated during D$_2$O electrolysis, which increases in presence of an applied magnetic field. The combined evidence suggests that electrochemical loading is associated with both charged-particle and slow neutron induced particle track signals, although the underlying mechanism remains unidentified. The protocol presented here offers a simple reference experiment that can be replicated and extended to investigate nuclear activity under eV-scale energy inputs. The results also highlight the need to consider this neutron activity when performing neutron-flux measurements in experiments involving deuteron-beam bombardment on electrochemical deuteron-loaded targets~\cite{chen2025electrochemical}.

In summary, a reproducible method has been established for isolating and quantifying low-intensity, slow-neutron induced particle track signal during D$_2$O electrolysis. These findings should motivate further studies to identify the physical mechanism and explore its dependence on material and electrolyte parameters.

\section{Conclusions}
   This study demonstrates reproducible detector signatures consistent with slow neutron capture like events observed during D$_2$O electrolysis involving D-Pd deposition on Pt cathode, using a differential CR-39–based detection methodology designed to discriminate neutron capture consistent tracks from charged particle and background contributions. Under an applied magnetic field of $0.25~\mathrm{T}$, the boric-acid coated BCR detector, consistently recorded a statistically significant average excess of $371 \pm 10$ tracks relative to an adjacently placed uncoated CR-39 (CCR) detector, which corresponds to the characteristic BCR $>$ CCR signature consistent with thermal-neutron capture on $^{10}$B nuclei within the present experimental constraints. Calibration of the BCR response indicates that the observed excess track density corresponds to a detector-equivalent slow-neutron flux estimate of approximately $\sim 6.7 \pm 0.2$~cm$^{-2}$~s$^{-1}$ at the detector position. Removal of the applied magnetic field resulted in a reduction of the detector response by a factor of $\sim$ 6, demonstrating a strong empirical dependence of the neutron like detector signal on the applied field. A critical negative control, consisting of light-water (H$_2$O) electrolysis performed under otherwise identical conditions, yielded track densities on BCR indistinguishable from CCR levels. Taken together, these observations establish a reproducible experimental correlation between D$_2$O electrolysis involving D-Pd deposition on a Pt cathode and the appearance of slow neutron like detector signatures, while excluding environmental radiation and detector artifacts as plausible explanations within the scope of the  present controls.
  
  The present results further indicate that electrochemically prepared metal deuterides can exhibit reproducible low level neutron-like detector signatures, underscoring the importance of careful background characterization and detector discrimination when interpreting neutron-related measurements in electrochemical environments and in broader contexts involving deuteron-loaded targets. While no theory is claimed to explain the observed effects, and no time-resolved information on neutron emission or detector response is accessible due to the passive, integrating nature of CR-39 detectors, the present work establishes a detector validated and control verified experimental protocol for resolving low intensity slow neutron signatures. Future investigations will focus on time-resolved neutron diagnostics, systematic mapping of magnetic field dependence, and moderator based energy selective measurements to further constrain the neutron energy spectrum and emission characteristics.

\nocite{*}
\bibliography{apssamp}


\clearpage
\onecolumngrid 
\begin{center}
	\textbf{\large Supplemental Material to “CR-39 track detector signatures of slow neutron like signals in Heavy-water electrolysis”
	\\
	\vspace{0.2em}
	by Ankit Kumar, Tushar Verma, Pankaj Jain, Raj Ganesh Pala and K.~P.~Rajeev }
\end{center}
\vspace{-0.5cm} 
\setcounter{equation}{0}
\setcounter{figure}{0}
\setcounter{table}{0}
\setcounter{section}{0}
\renewcommand{\theequation}{S\arabic{equation}}
\renewcommand{\thefigure}{S\arabic{figure}}
\renewcommand{\thetable}{S\arabic{table}}

\vspace{0.8cm}
\maketitle
\twocolumngrid
The detailed experimental methodology of this study is provided in Section~I. The raw data involved in this study are provided in Section~II. Figures, equations, and tables appearing in this Supplemental Material are labeled S1, S2, \dots, to distinguish them from those in the main article.

\section{Experimental Method} 
\subsection{Materials}

\textbf{Particle detector:} Columbia Resin-39 chips of size $6 $~mm$\times 4$~mm$ \times 1.5$~mm (Tastrak, Bristol, UK), $42~\mu$m thick polypropylene cello tape, $20~\mu$m thick polyethylene sheet, boric acid powder (Fischer Scientific), cyanoacrylate adhesive (Pidilite) and 15~mL polypropylene centrifuge tube (Tarsons).

\textbf{Electrochemical setup:} D$_2$O ($99.82$ to $99.91$ mass \% (nuclear grade purity) from Heavy Water Board, India, PdCl$_2$ (Sigma Aldrich), LiCl (Sigma Aldrich), platinum wire ($0.5$~mm thick and $6$~cm length, Alfa Aesar), graphite rod ($6$~mm thick, $6$~cm length, Fischer Scientific), borosilicate beaker of $25$~mL 
($ID \times OD \times H \approx 32.6$~mm$ \times 34$~mm$ \times 50$~mm, Borosil), Teflon plastic sheet ($40$~mm$ \times 40$~mm$ \times 1.5$~mm), two neodymium magnets ($5$~cm$ \times 2.5$~cm$ \times 5$~cm).

\textbf{Power supply:} Constant current DC supply (HP 177).

\textbf{Etching setup and reagents:} Borosilicate beaker of $25$~mL ($ID \times OD \times H \approx 32.6$~mm$ \times 34$~mm$ \times 50$~mm, Borosil) and $250$~mL ($ID \times OD \times H \approx 68$~mm$ \times 70$~mm$ \times 95$~mm, Borosil), glass tube ($ID \times OD \times H \approx 13.8$~mm$ \times 	24$~mm$ \times 40$~mm, Borosil), NaOH (Loba Chemie), deionized light water (H$_2$O), hot plate, mercury thermometer.

\textbf{Imaging :} Leica DM 2700P microscope.

\subsection{Etching Procedure}

We first present the general procedure to reveal particle tracks from any irradiated or unexposed CR-39 sample. All CR-39 samples were positioned in the gap formed between a hollow glass tube ($24$~mm $\times$ $40$~mm) placed concentrically inside a $25$~mL capacity beaker containing $20$~mL of preheated $6$~M NaOH souution in deionized H$_2$O, at $70 \pm 2~^\circ$C for 6~h [see Fig.~\ref{fig:s1}(2)]. The $25$~mL beaker was placed in a $70~^\circ$C water bath held inside a $250$~mL beaker, which was placed on a hot plate. After etching, samples were rinsed under tap water, gently blotted dry, and immediately sealed in centrifuge tube, which took about 2 minutes (resulting in CR-39 exposed to the ambient radiation of the laboratory for $\sim 2$~min). Any etching of CR-39 samples (unexposed or exposed to radiation) in this study follows the same procedure as discussed above.

\subsection{Microscopic Imaging and track counting of Etched CR-39}

Each CR-39 sample after etching was immediately imaged using an optical microscope across four areal regions, one at each corner, with each measuring $4.8$~mm$^2$ ($2.5$~mm$ \times 1.9$~mm) [see Fig.~\ref{fig:s1}(3)], yielding four track measurements per CR-39 sample. Since all images cover the same area throughout the study, particle tracks per $4.8$~mm$^2$ are hereafter referred to simply as particle tracks. Oval particle tracks with a semi-minor axis greater than $7~\mu$m were only counted as a track see Sec.~I~F of the Supplemental Material. We note that imaging each CR-39 required $\sim 2$~minutes (hence during imaging, the CR-39 is exposed to laboratory ambient radiation once again for $\sim 2$~min, after etching). Track counting on any CR-39 sample (unexposed or exposed to radiation) in this study follows the same procedure as discussed above.

\subsection{Baseline Track Measurements of CR-39}

Ten unexposed CR-39 samples, each of size $6 $~mm$\times 4$~mm$ \times 1.5$~mm [see image of bare CR-39 (A) in Fig.~\ref{fig:s1}(1)], as obtained from the supplier packed in $50~\mu$m aluminized Mylar pouch, were etched and imaged to obtain information about the pre-existing tracks. Consequently, for the ten etched samples, forty regions were imaged and the corresponding particle-track counts were obtained and averaged to determine the baseline particle density, which was subtracted from all subsequent track measurements taken using this CR-39 to obtain the net particle tracks recorded during any exposure.

\subsection{Detector Fabrication, Electrolysis Procedures, Background Measurements of BCR and CCR Detector}

\begin{figure*}[!htbp]
	\centering
	\includegraphics[width=\textwidth]{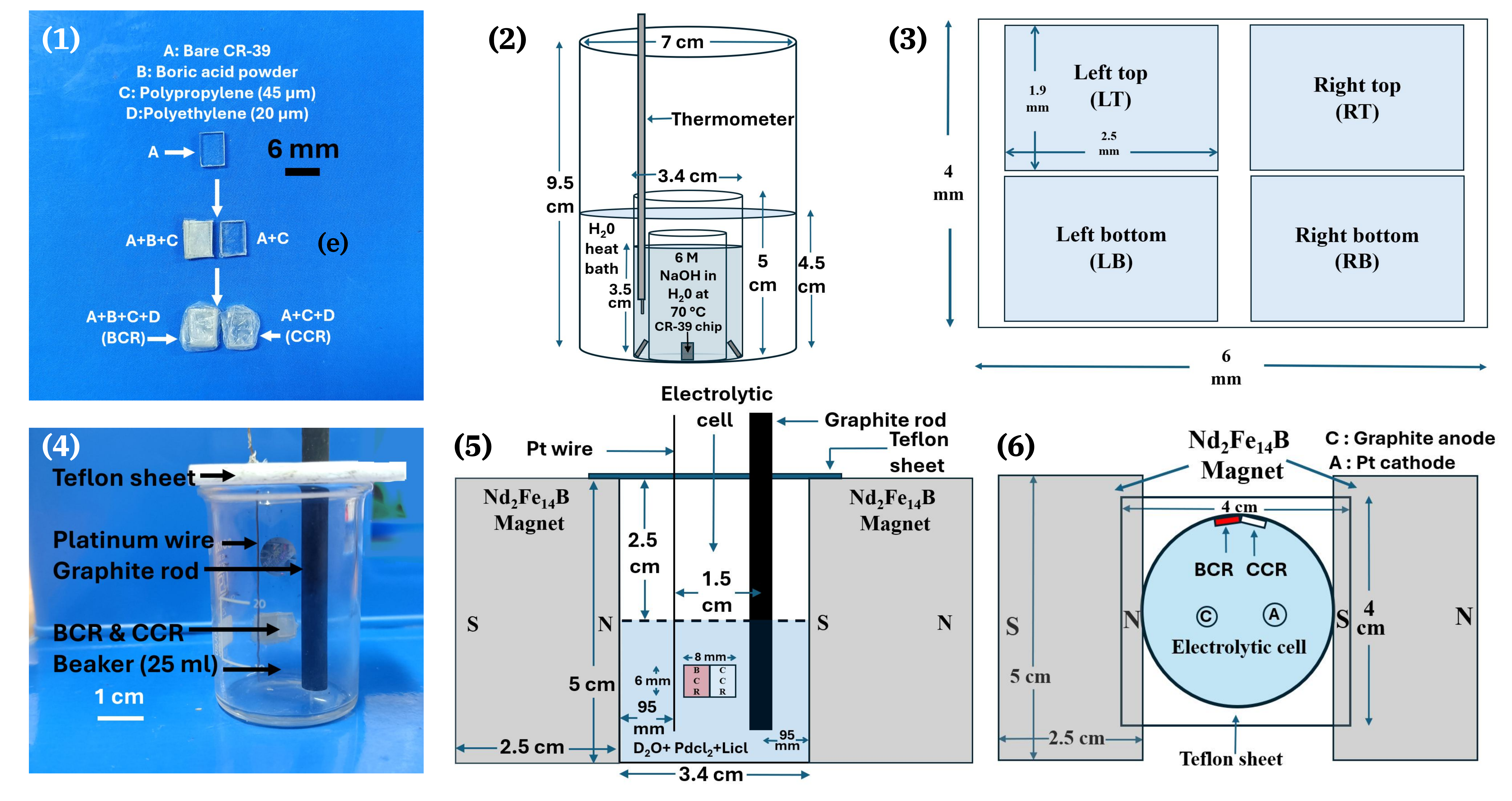}
	\caption{\justifying (1) Preparation, etching, imaging, and experimental configuration of CR-39 detectors.
		(1) Detector preparation. A bare CR-39 sheet (A) is covered with a polypropylene tape (C) to form the CCR. For the BCR, a boron layer (B) is deposited on the polypropylene, which is then placed on the bare CR-39, yielding the configuration A+B+C. Each assembly is further sealed in a $20~\mu$m polyethylene cover (D), resulting in final CCR (A+C+D) and BCR (A+B+C+D) detectors. The sole difference between the two detectors is the presence of the boron layer in the BCR. (2) Illustration of CR-39 etching: A $250$-mL beaker ($9.5$~cm $\times$ $7$~cm) filled with water is placed on a hot plate and maintained at $70~^\circ$C to serve as the water bath. A $25$-mL beaker containing the etching solution is positioned within the water bath. The CR-39 samples are held vertically by securing them between the wall of the $25$-mL beaker and a hollow glass cylinder placed concentrically inside it. (3) Imaging and track-counting scheme for an etched CR-39 sample: Four regions, left top (LT), right top (RT), right bottom (RB) and left bottom (LB), each measuring $2.5$~mm $\times$ $1.9$~mm (each $4.8$~mm$^2$ area), were imaged on the etched CR-39 detector. Particle tracks within these four regions were counted. (4) Cathode--anode assembly and detector placement within the electrochemical cell: The cathode and anode are mounted perpendicular to the Teflon sheet that covers the electrolytic cell as a lid, beneath which the BCR and CCR detectors are positioned for particle-track recording during electrolysis. Schematic of experimental assembly, (5) side view and (6) top view.}
	\label{fig:s1}
\end{figure*}

To generate the external magnetic field, two cuboidal neodymium magnets, were mounted on the laboratory bench with their $5$~cm$\times 2.5$~cm faces resting on the surface. The $5$~cm $\times 5$~cm faces were positioned parallel in an attracting configuration, separated by a $3.4$~cm gap, to accommodate electrolytic cell [see Fig.~\ref{fig:s1}(5) and \ref{fig:s1}(6)].

A solution of $1.25$~mM PdCl$_2$ and $150$~mM LiCl in $20$~mL of heavy water (D$_2$O) or light water (H$_2$O) was prepared for use as the electrolyte. The electrodes used are a graphite rod ($6$~mm diameter, $6$~cm length) as the anode and a platinum wire ($0.5$~mm diameter, $6$~cm length) as the cathode. The electrodes were mounted parallel to each other and perpendicular to a square Teflon sheet ($4 \times 4$~cm$^2$, $1.5$~mm thick), spaced $1.5$~cm apart. Each electrode extended $\sim 4.5$~cm on one side of the sheet and the remainder on the other, positioned collinearly and equidistant from the sheet’s center [see Fig.~\ref{fig:s1}(4)--\ref{fig:s1}(6)].

We now describe the preparation of BCR and CCR detectors, which were prepared using an identical procedure, differing only in the presence of the boron coating. For BCR preparation, boric acid was spread over a
$10~\mathrm{mm} \times 10~\mathrm{mm}$ adhesive region of a
$42~\mu\mathrm{m}$-thick polypropylene tape. After removal of excess material
by gentle air blowing, a visibly uniform boric acid layer remained adhered to the tape. To quantify the typical variation in the mass of boric acid coated on
polypropylene sheets ($10~\mathrm{mm} \times 10~\mathrm{mm}$ each), the masses
of ten tapes were measured before and after coating.The resulting coating masses were $1.4 \pm 0.1$, $1.2 \pm 0.1$, $1.3 \pm 0.1$, $1.3 \pm 0.1$, $1.4 \pm 0.1$, $1.3 \pm 0.1$, $1.3 \pm 0.1$, $1.4 \pm 0.1$, $1.4 \pm 0.1$, and $1.3 \pm 0.1~\mathrm{mg}$, with a mean of $1.3~\mathrm{mg}$ and a standard deviation of $0.1~\mathrm{mg}$,
indicating a small sample-to-sample variation. A $6~\mathrm{mm} \times 4~\mathrm{mm}$ piece was cut from such boric acid coated polypropylene sheet and affixed onto a bare CR-39 detector chip such that the boron layer is sandwiched between the polypropylene pouch and the CR-39 substrate. The resulting assembly was then sealed in a $20~\mu$m thick waterproof polyethylene pouch, resulting in BCR with a total effective plastic thickness of $65~\mu$m on the boron-coated/neutron detection surface. The CCR was prepared in the same manner but without boric acid coating, so the only difference between the BCR and CCR is the presence of the boron coating on the BCR [see Fig.~\ref{fig:s1}(1)]. Detector preparation was completed within two minutes, during which the CR-39 was exposed to the laboratory for only $\sim 0.5$~min. The BCR and CCR were immediately mounted side by side on the inner curved surface of a $25$~mL cylindrical borosilicate beaker ($30$~mm diameter, $50$~mm height) using a cyanoacrylae adhesive. They spanned a length of $1$--$1.8$~cm from the base of the cell, with their detecting surfaces facing towards electrodes [Fig.~\ref{fig:s1}(4) and \ref{fig:s1}(5)].

The beaker was then filled with $20$~mL of electrolyte, reaching a height of $2.5$~cm from the bottom. The Teflon sheet holding the electrodes was placed on the beaker so that $4.5$~cm of each electrode was suspended vertically, with approximately $2.2$~cm immersed in the electrolyte [Fig.~\ref{fig:s1}(5) and \ref{fig:s1}(6)]. The Teflon sheet was placed as a cover over the beaker, with its center aligned along the beaker’s longitudinal axis, and the electrodes positioned to remain equidistant from the line of contact between the detectors on the inner curved surface of the beaker [Fig.~\ref{fig:s1}(5) and \ref{fig:s1}(6)].

For electrolysis in the presence of a magnetic field, the cell was placed in between the two fixed neodymium magnets, such that the magnetic field was directed from the cathode to the anode and measured $0.25$~T at the center of the beaker [Fig.~\ref{fig:s1}(5) and \ref{fig:s1}(6)]. For experiments without a magnetic field, electrolysis was conducted in the absence of the magnet assembly. In either case, the portions of the electrodes extending above the Teflon cap were connected to a constant current source, and the cells were ready for electrolysis. We note that the process starting from detector fabrication to making the cell ready for electrolysis typically was completed within $5$~minutes, during which the CR-39 samples remained in their bare form only for about half a minute (during detector fabrication).

The resulting heavy-water-based experimental assembly, in the presence of a magnetic field, was further used in two configurations for recording particle tracks on BCR and CCR: an unpowered state (electrolysis-OFF, for background track estimation) and a powered state (for electrolysis-ON track estimation).

Five unpowered D$_2$O-based electrolytic assemblies in the presence of a magnetic field were utilized to assess background track accumulation on BCR and CCR detectors for a total duration of 16 weeks. Following exposure, the CR-39 substrates were chemically etched and analyzed by optical imaging, yielding 20 images for each detector type. Baseline track values were subtracted from the mean of the 20 measured values for each detector, and the resulting net counts were normalized by the exposure time to obtain the weekly background tracks for the BCR and CCR, respectively.

For electrolysis track-density estimation, the cell was powered up continuously for 7 days, during which, for the first 3 days, a constant current of $50.0 \pm 0.1~\mu$A was applied, followed by an increased current of $100 \pm 5$~mA at $7.0 \pm 0.5$~V for the remainder of the time. To compensate for electrolyte loss due to gas evolution, the electrolyte volume was constantly maintained by adding D$_2$O/H$_2$O to the electrochemical cell on a daily basis. At the conclusion of electrolysis, the underlying CR-39 substrates were etched, imaged, and optically analyzed to obtain track counts on them as described earlier. A total of five D$_2$O electrolysis experimental runs were performed in the presence of an applied $0.25$~T magnetic field. Additionally, two control configurations including a D$_2$O electrolysis in zero applied magnetic field and an H$_2$O electrolysis under $0.25$~T magnetic field were also investigated.

\onecolumngrid
\vspace{-0.5cm}

\subsection{Track counting illustration}
\begin{center}
	\includegraphics[width=0.25\textwidth]{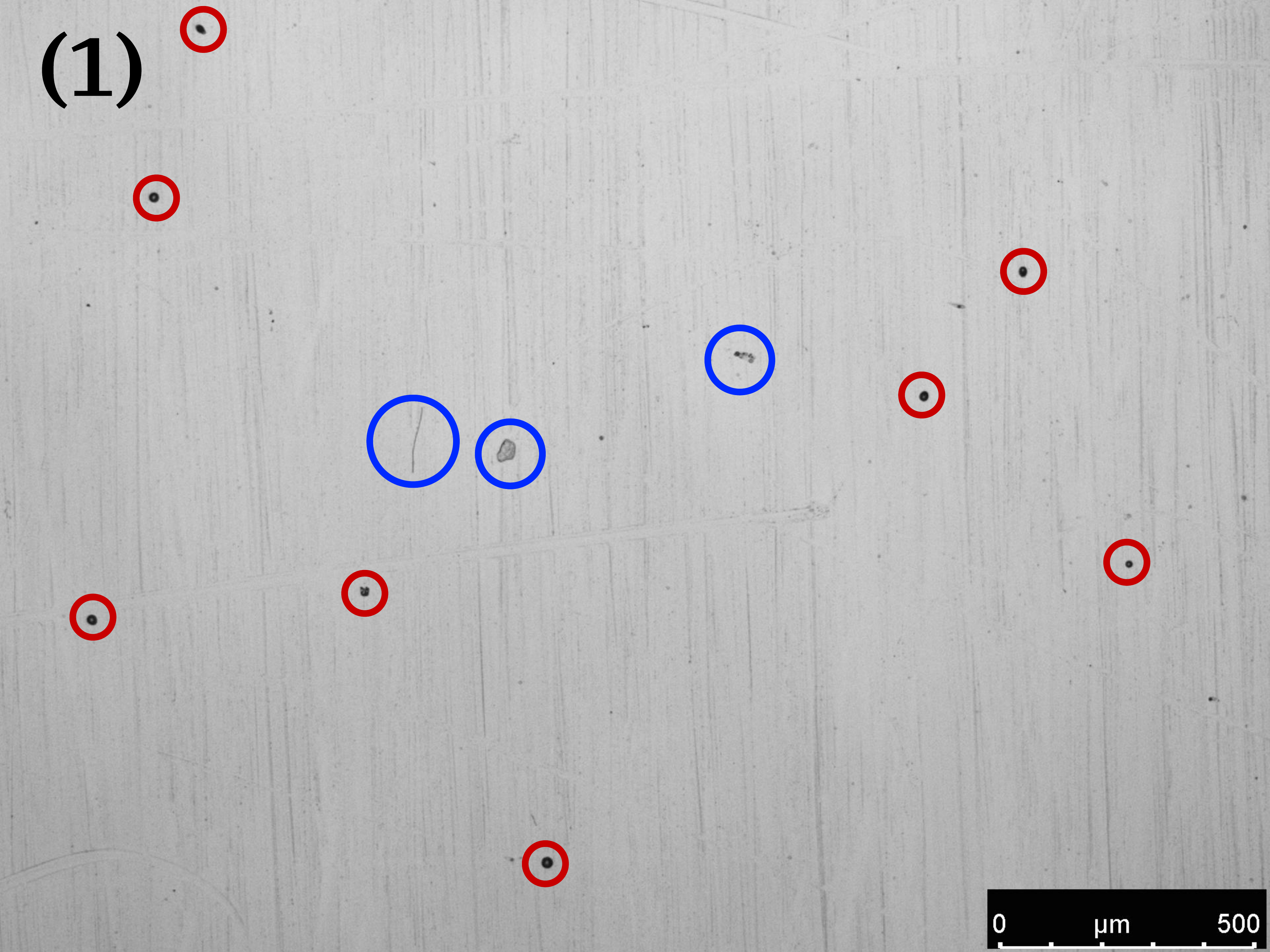}
	\includegraphics[width=0.25\textwidth]{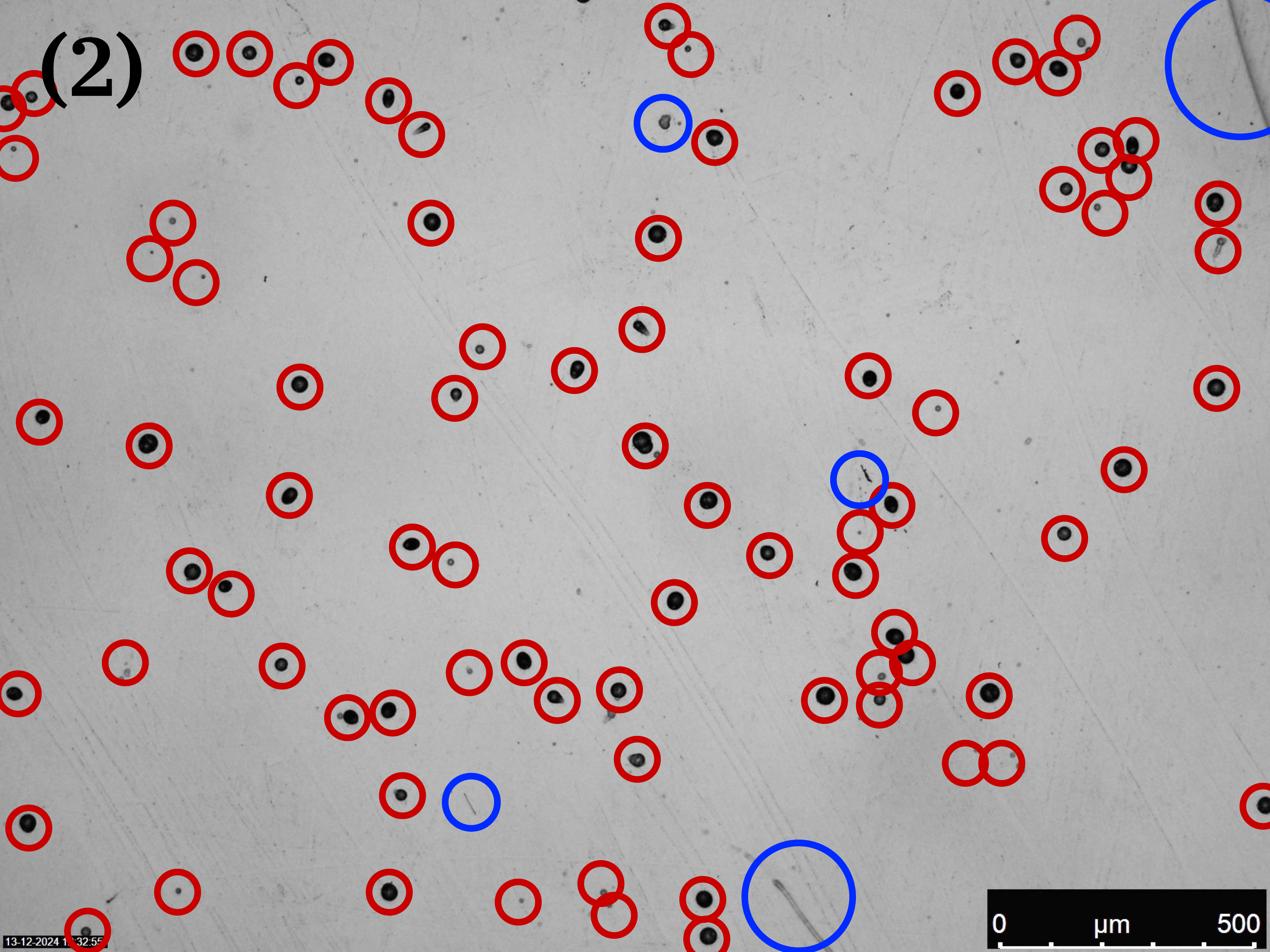}
	\includegraphics[width=0.25\textwidth]{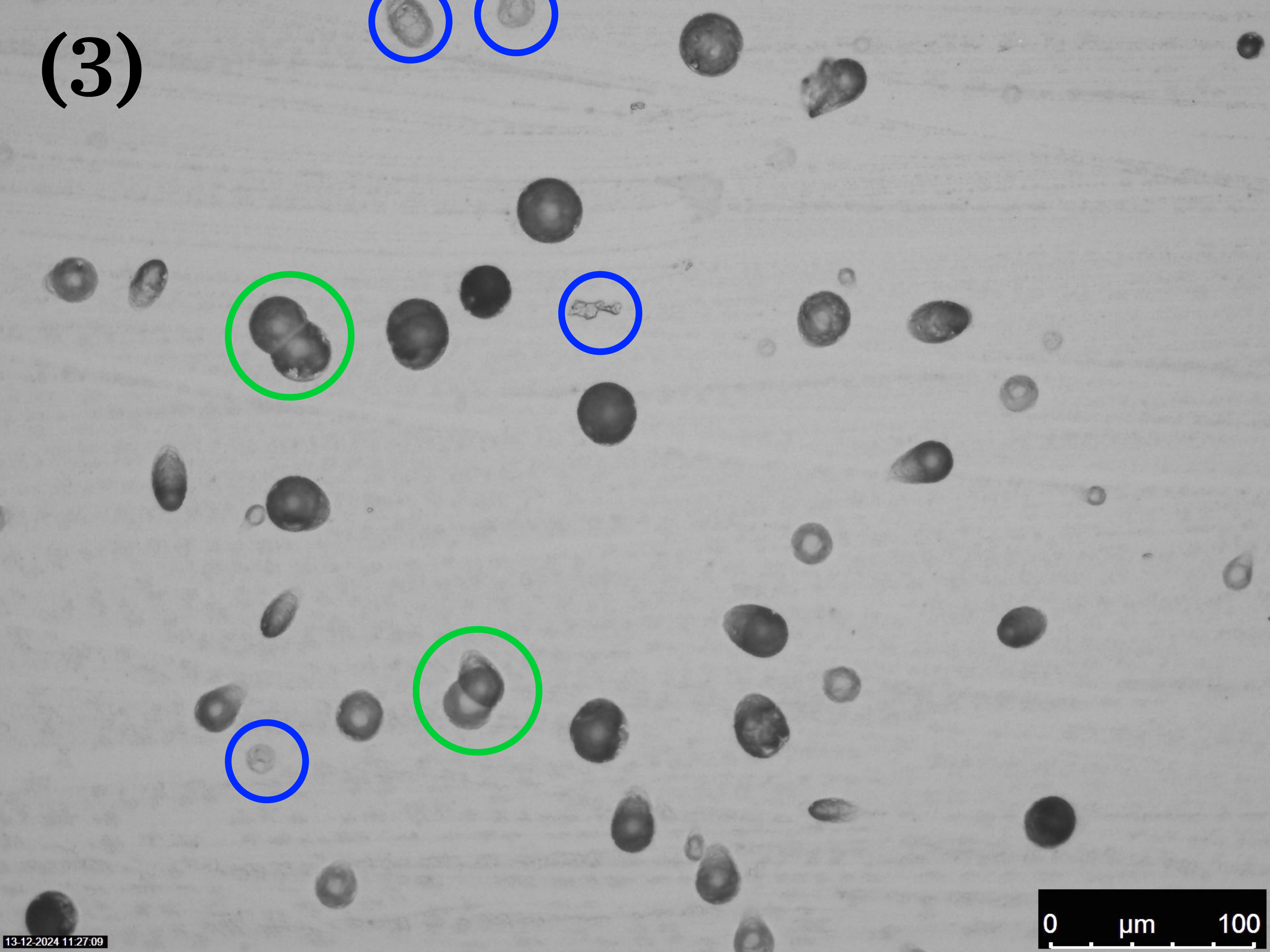}
	
	\captionsetup{type=figure, width=\linewidth, justification=justified, format=plain}
	\captionof{figure}{\justifying Illustration of track counting on CR-39 images. Features encircled in red correspond to structures identified and counted as individual particle tracks. Features encircled in green represent clustered or non-isolated tracks that are attributed to a single particle event and are therefore counted as one track in this study. Features encircled in blue indicate CR-39 defects or artifacts and are excluded from the track count.
		(1) Track counting on a 1.8~mm $\times$ 2.5~mm unexposed, etched CR-39 detector, showing 8 tracks.
		(2) Track counting on a 1.8~mm $\times$ 2.5~mm electrolysis-exposed, etched CR-39 detector, exhibiting 81 tracks, corresponding to a substantially higher particle-track density than that of the unexposed chip.
		(3) $5$X $(\sim 621~\mu m \times 466~\mu m)$ image of CR-39 detector, illustrating the geometrical features of particle tracks. Double and triple tracks are encircled in green.}
	\label{fig:s2}
\end{center}

\onecolumngrid
\section{Raw Data}
\vspace{-0.6cm}
\subsection{Baseline Measurement Data}
\begin{center}
	\includegraphics[width=\textwidth]{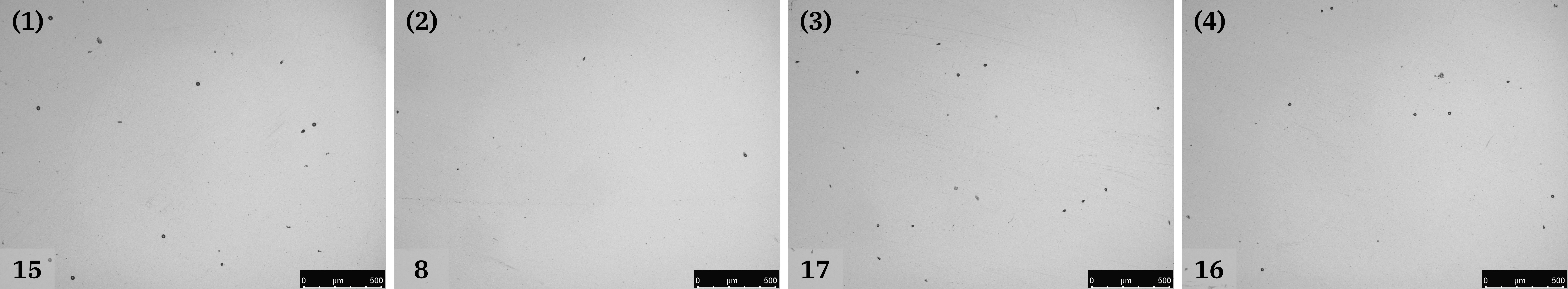}\\[0.2em]
	\includegraphics[width=\textwidth]{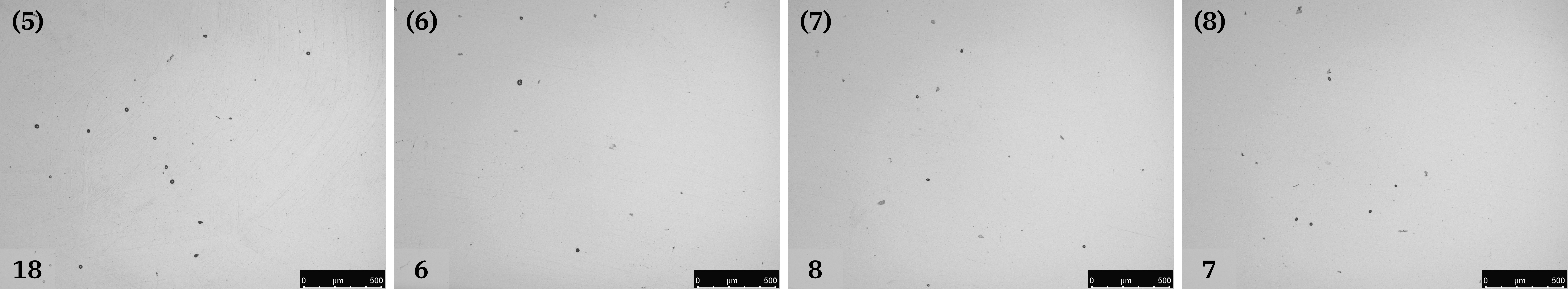}\\[0.2em]
	\includegraphics[width=\textwidth]{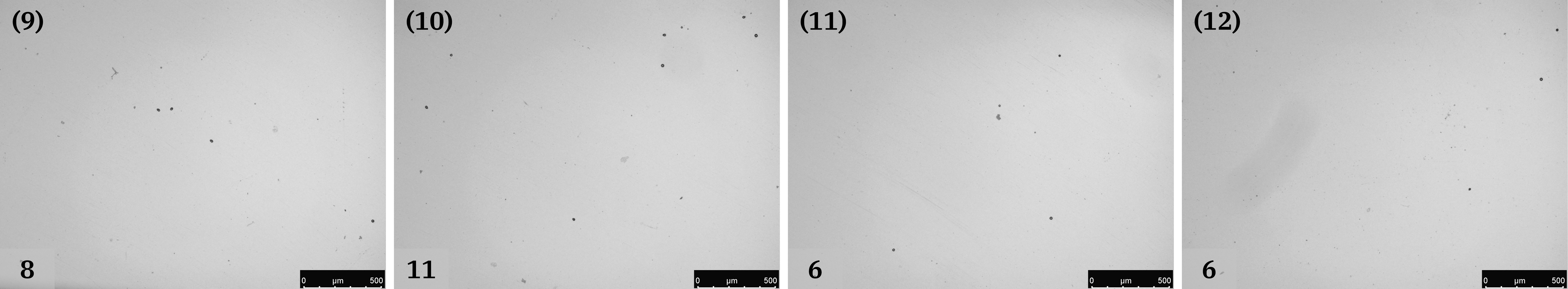}\\[0.2em]
	\includegraphics[width=\textwidth]{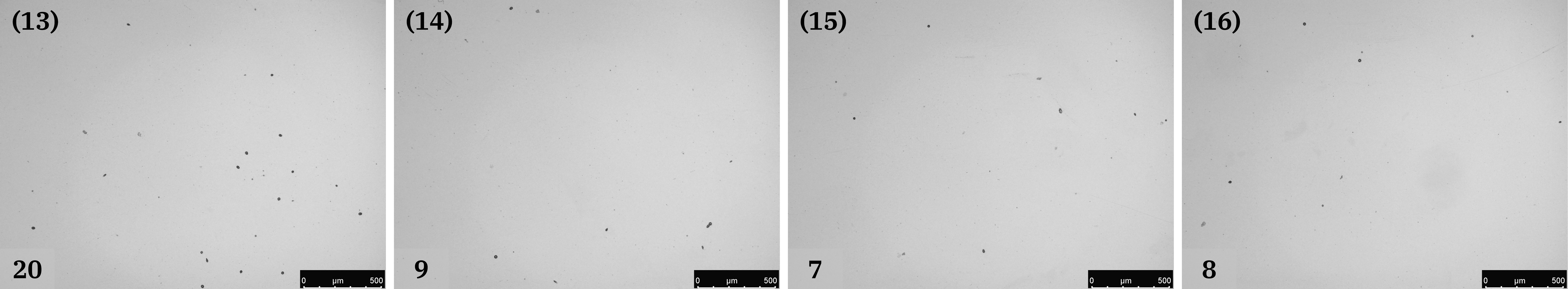}\\[0.2em]
	\includegraphics[width=\textwidth]{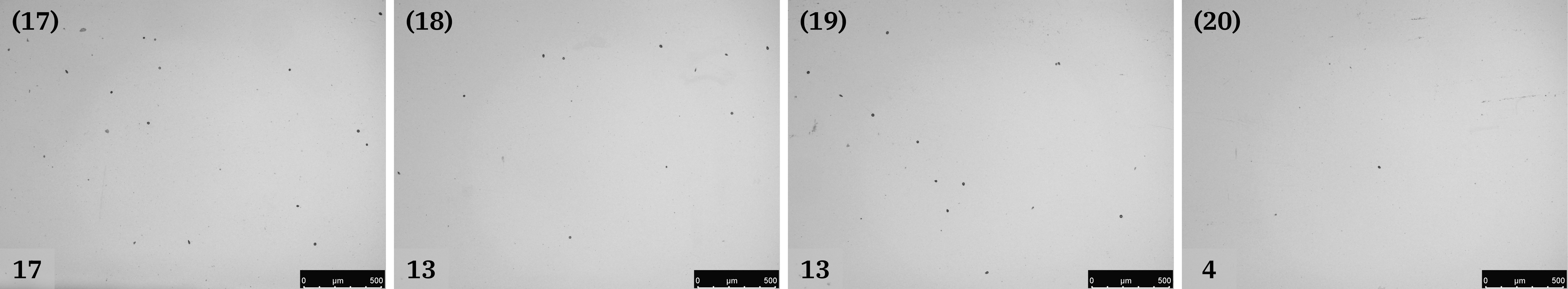}\\[0.2em]
	\includegraphics[width=\textwidth]{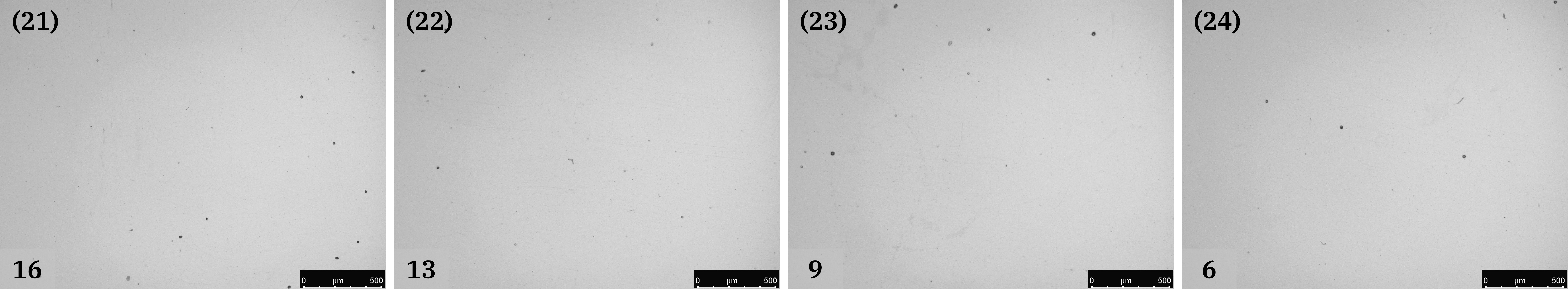}\\[0.2em]
	\includegraphics[width=\textwidth]{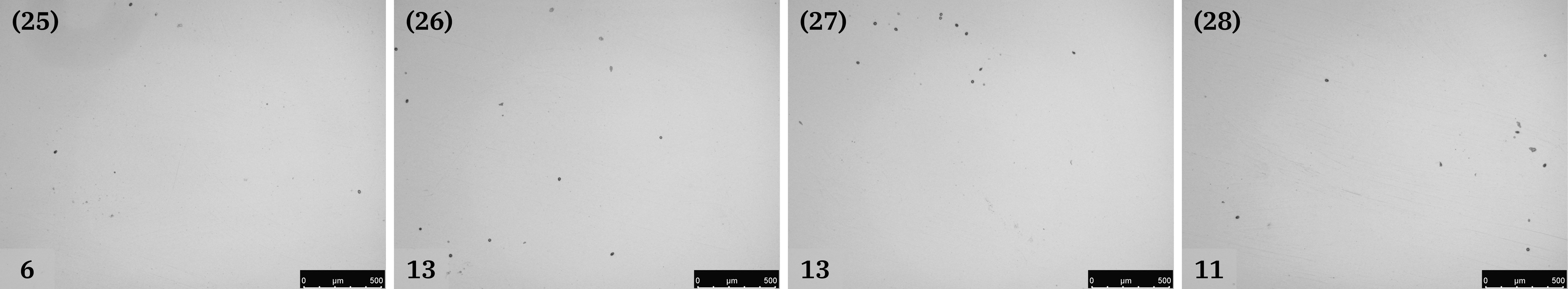}\\[0.2em]
	\includegraphics[width=\textwidth]{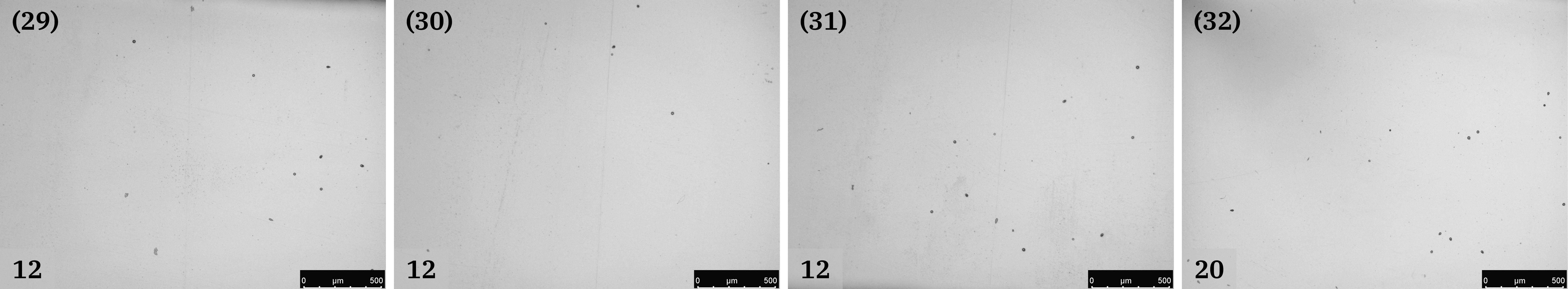}\\[0.2em]
	\includegraphics[width=\textwidth]{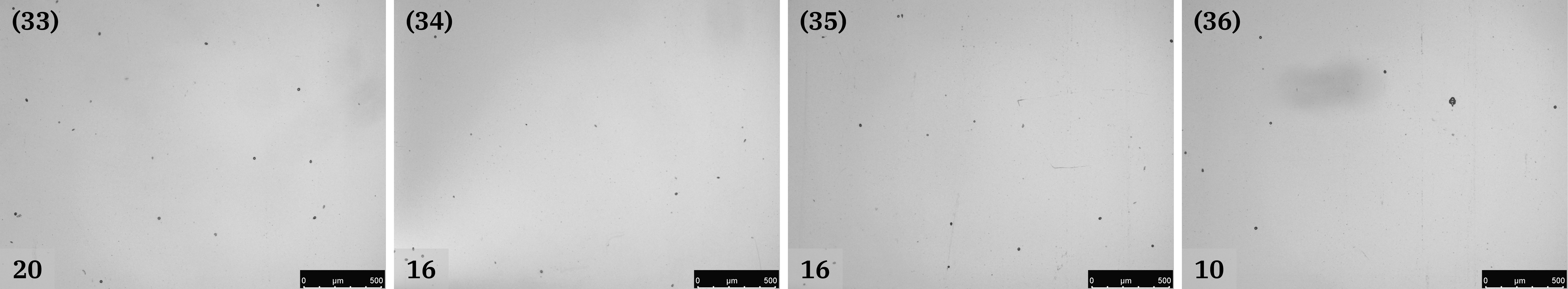}\\[0.2em]
	\includegraphics[width=\textwidth]{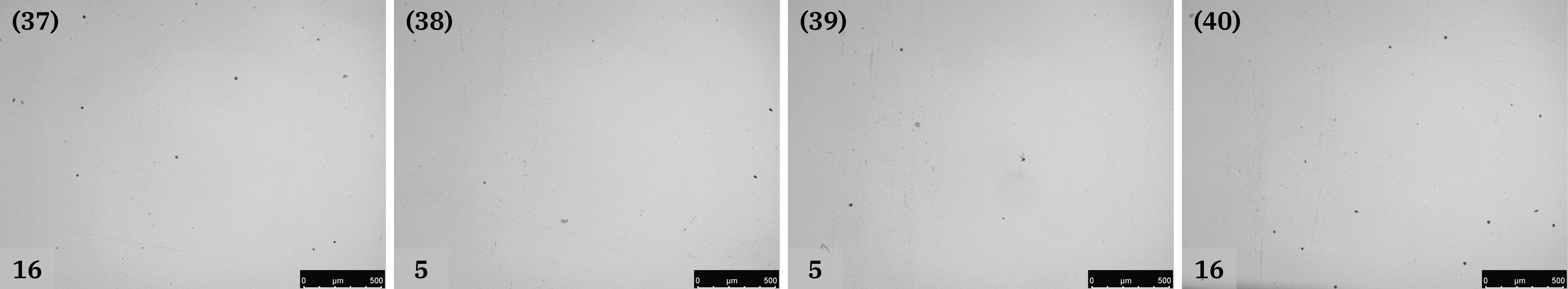}
	
	\captionsetup{type=figure, width=\linewidth, justification=justified, format=plain}
	\captionof{figure}{\justifying Optical micrographs of ten unexposed CR-39 detector samples, used to determine  baseline tracks. Each row of image, from left to right represents Left top (LT), right top (RT), right bottom (RB), and left bottom (LB) region of a sample, [sample~1: 1--4; sample~2: 5--8; sample~3: 9--12; sample~4: 13--16; sample~5: 17--20; sample~6: 21--24; sample~7: 25--28; sample~8: 29--32; sample~9: 33--36; sample~10: 37--40]. The track count for each image is indicated at its lower left corner. The mean and standard error of the 40 counts obtained on these 40 images were determined to be \(11.6 \pm 0.7\) per \(4.8~\mathrm{mm}^2\) as the baseline track count value of CR-39, to be used in all calculations in this study.}
	\label{fig:s3}
\end{center}

\subsection{BCR Background Measurement Data}

\begin{center}
	\includegraphics[width=\textwidth]{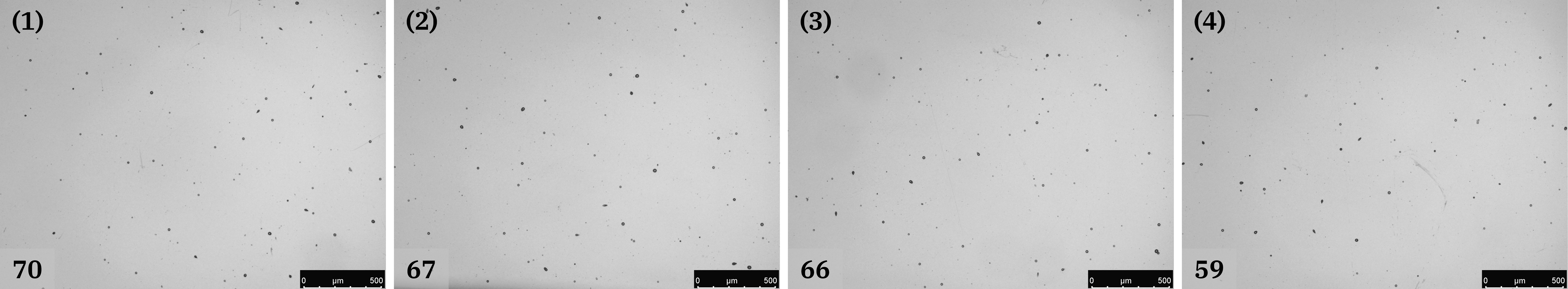}\\[0.2em]
	\includegraphics[width=\textwidth]{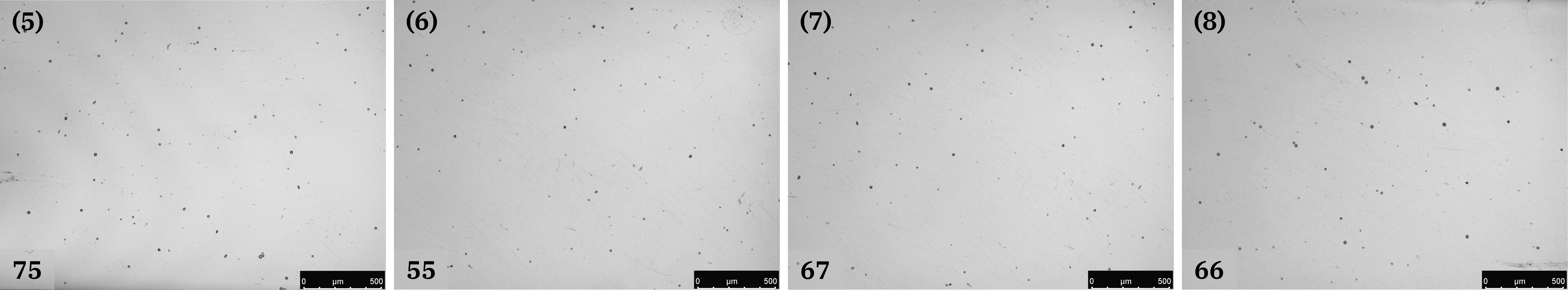}\\[0.2em]
	\includegraphics[width=\textwidth]{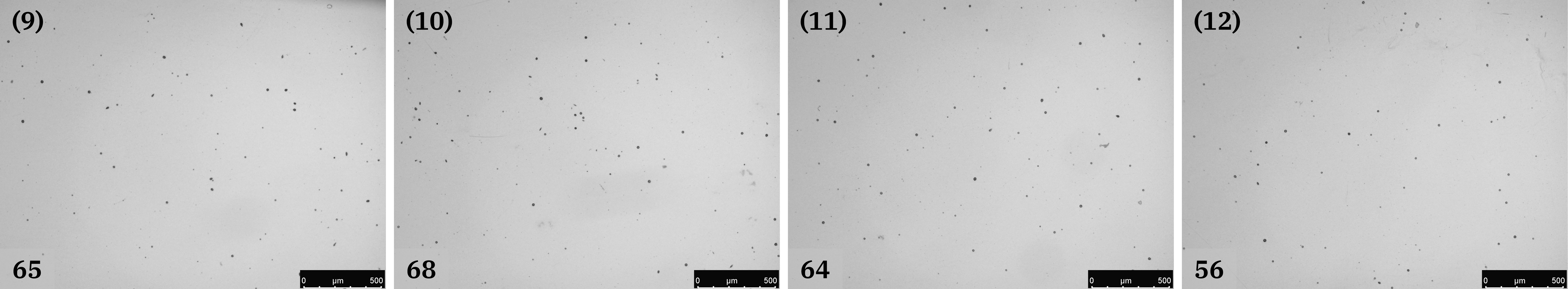}\\[0.2em]
	\includegraphics[width=\textwidth]{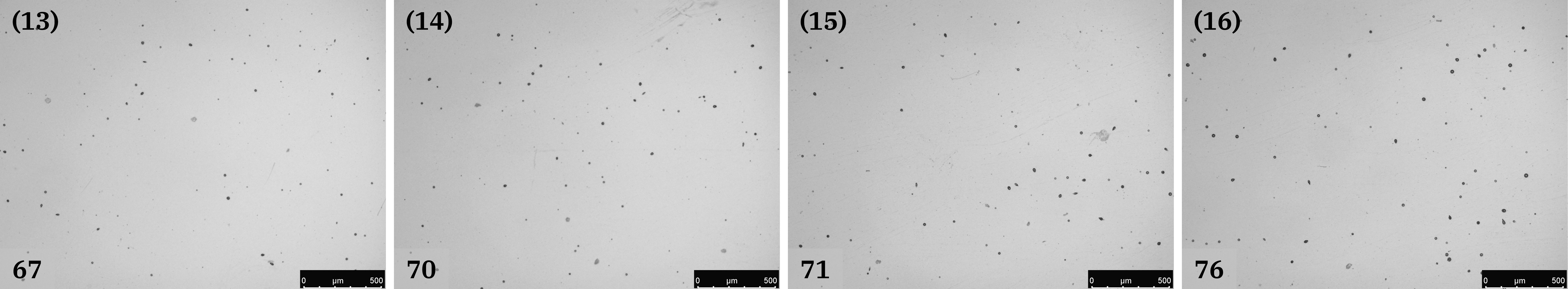}\\[0.2em]
	\includegraphics[width=\textwidth]{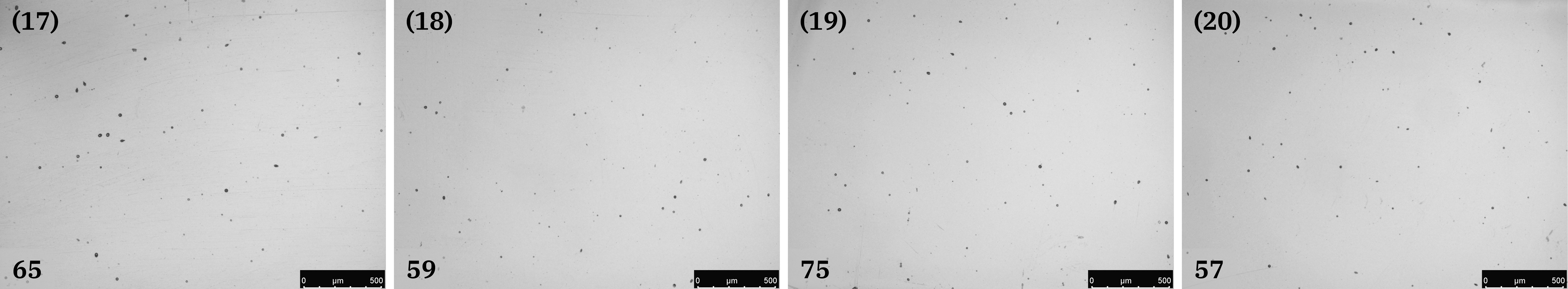}

	\captionsetup{type=figure, width=\linewidth, justification=justified, format=plain}
	\captionof{figure}{\justifying Optical micrographs of five etched BCR, after being placed inside five independent unpowered electrolytic cells, each subjected to a static magnetic field of 0.25~T for 16-week period. Each row of image, from left to right represents LT, RT, RB, and LB region, of a BCR sample [BCR-1: 1--4; BCR-2: 5--8; BCR-3: 9--12; BCR-4: 13--16; BCR-5: 17--20]. The mean and standard error of the 20 counts obtained on these 20 BCR images were determined to be \(65.9 \pm 1.4\)per \(4.8~\mathrm{mm}^2\). The baseline track counts is subtracted from the obtained mean value resulting in \(54.3 \pm 1.6\)per \(4.8~\mathrm{mm}^2\), which yields \(3.39 \pm 0.10\) tracks per \(4.8~\mathrm{mm}^2\) per week, as the BCR background value to be used in all calculations in this study.}
	\label{fig:s4}
\end{center}

\subsection{CCR Background Measurement Data}

\begin{center}
	\includegraphics[width=\textwidth]{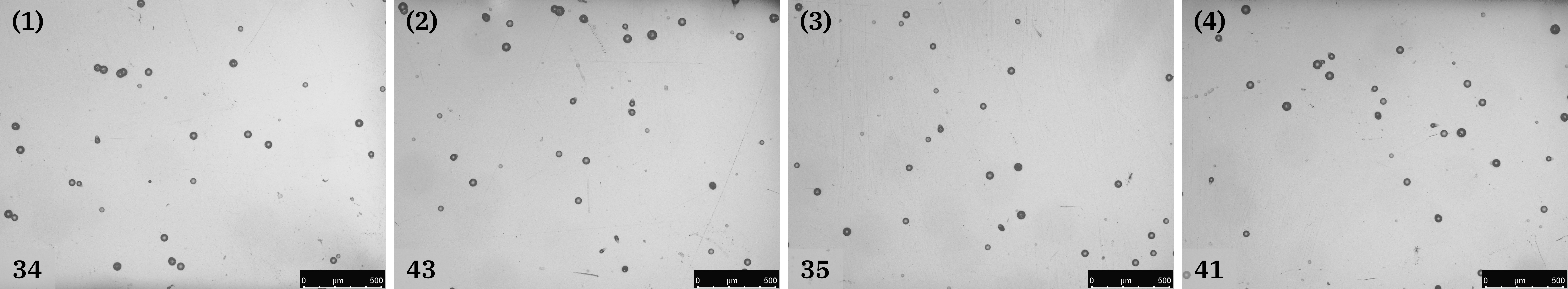}\\[0.2em]
	\includegraphics[width=\textwidth]{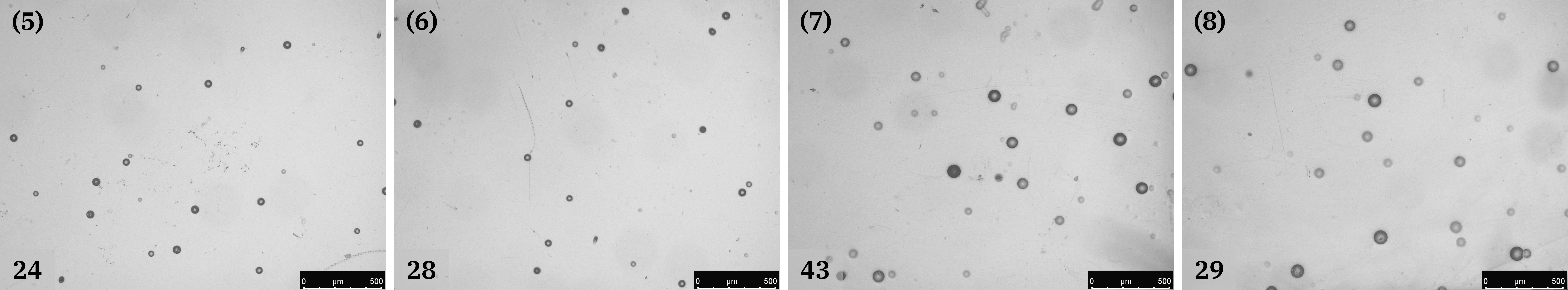}\\[0.2em]
	\includegraphics[width=\textwidth]{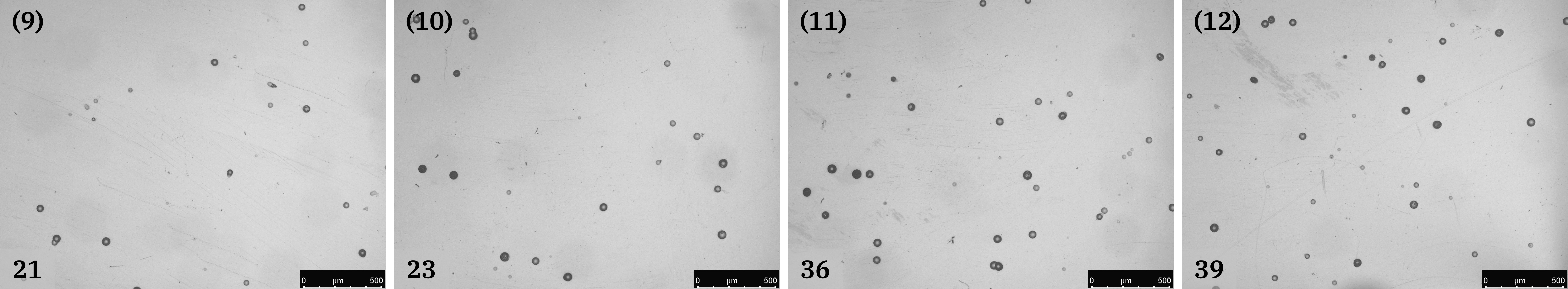}\\[0.2em]
	\includegraphics[width=\textwidth]{CR39/s4.4.pdf}\\[0.2em]
	\includegraphics[width=\textwidth]{CR39/s4.5.pdf}
	
	\captionsetup{type=figure, width=\linewidth, justification=justified, format=plain}
	\captionof{figure}{\justifying Optical micrographs of five etched CCR, after being placed inside five independent unpowered electrolytic cells, each subjected to a static magnetic field of 0.25~T for 16-week period. Each row of image, from left to right represents LT, RT, RB, and LB region, of a CCR sample [CCR-1: 1--4; CCR-2: 5--8; CCR-3: 9--12; CCR-4: 13--16; CCR-5: 17--20].The mean and standard error of the 20 counts obtained on these 20 CCR images were determined to be \(31.6 \pm 1.5\)per \(4.8~\mathrm{mm}^2\). The baseline track counts is subtracted from the obtained mean value resulting in \(20.0 \pm 1.7\)per \(4.8~\mathrm{mm}^2\), which yields \(1.25 \pm 0.11\) tracks per \(4.8~\mathrm{mm}^2\) per week, as the CCR background value to be used in all calculations in this study.}
	\label{fig:s5}
\end{center}

\subsection{BCR and CCR Track Distributions Recorded during multiple D$_2$O Electrolysis runs under an Applied 0.25 T Magnetic Field and Their Statistical Comparison}

\begin{center}
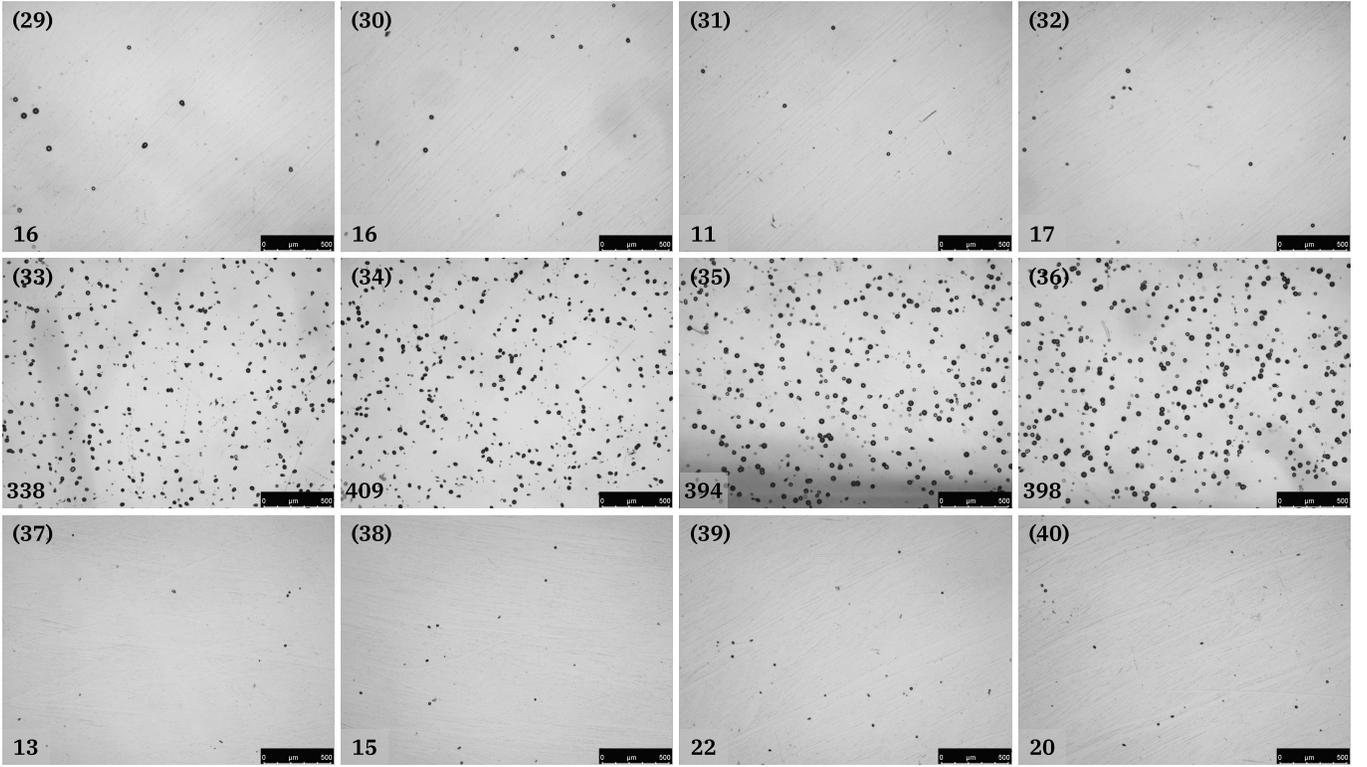

	\includegraphics[width=\textwidth]{CR39/s5.1.pdf}\\[0.2em]
	\includegraphics[width=\textwidth]{CR39/s5.2.pdf}\\[0.2em]
	\includegraphics[width=\textwidth]{CR39/s5.3.pdf}\\[0.2em]
	\includegraphics[width=\textwidth]{CR39/s5.4.pdf}\\[0.2em]
	\includegraphics[width=\textwidth]{CR39/s5.5.pdf}\\[0.2em]
	\includegraphics[width=\textwidth]{CR39/s5.6.pdf}\\[0.2em]
	\includegraphics[width=\textwidth]{CR39/s5.7.pdf}\\[0.2em]
	\includegraphics[width=\textwidth]{CR39/s5.8.pdf}\\[0.2em]
	\includegraphics[width=\textwidth]{CR39/s5.9.pdf}\\[0.2em]
	\includegraphics[width=\textwidth]{CR39/s5.10.pdf}
	
	\captionsetup{type=figure, width=\linewidth, justification=justified, format=plain}
	\captionof{figure}{\justifying Optical micrographs of BCR and CCR detectors from five independent D$_2$O electrolysis run performed under a $0.25$~T field for a week. Each row of image, from left to right represents LT, RT, RB, and LB regions, of any BCR or CCR sample. The track count for each image is indicated at the its lower left corner. Image labels correspond to: Trial 1—BCR (1–4), CCR (5–8); Trial 2—BCR (9–12), CCR (13–16); Trial 3—BCR (17–20), CCR (21–24); Trial 4—BCR (25–28), CCR (29–32); Trial 5—BCR (33–36), CCR (37–40). The mean track counts obtained by averaging the four imaged regions for each BCR and CCR in every experimental run were used to construct Fig.~\ref{fig:3} in the main text and are summarized in Table~\ref{tab:s1} below.}
	\label{fig:s6}
\end{center}

\begin{table*}[!h]
	
	\renewcommand{\arraystretch}{1.2} 
	\setlength{\tabcolsep}{8pt}      
	
	\caption{\justifying Mean track counts measured on BCR and CCR detectors for five independent D$_2$O electrolysis runs performed under an applied magnetic field of $0.25$~T as shown in Fig.~\ref{fig:s6} above. The values in 'mean tracks' column of this table was used to generate Fig.~\ref{fig:3} of the main text.}
	\label{tab:s1}
	\begin{ruledtabular}
		\begin{tabular}{cccccccc}
			Trial & Detector & LT & RT & RB & LB & Mean tracks \\
			\hline
			1 & BCR & 384 & 433 & 455 & 441 & $428 \pm 13$ \\
			& CCR & 18 & 16 & 19 & 21 & $18.5 \pm 0.9$ \\
			2 & BCR & 381 & 346 & 355 & 431 & $378 \pm 17$ \\
			& CCR & 12 & 19 & 18 & 13 & $15.5 \pm 1.5$ \\
			3 & BCR & 383 & 417 & 341 & 401 & $386 \pm 14$ \\
			& CCR & 15 & 15 & 16 & 16 & $15.5 \pm 0.3$ \\
			4 & BCR & 343 & 381 & 362 & 337 & $356 \pm 9$ \\
			& CCR & 16 & 16 & 11 & 17 & $15.0 \pm 1.2$ \\
			5 & BCR & 338 & 409 & 394 & 398 & $385 \pm 14$ \\
			& CCR & 13 & 15 & 22 & 20 & $17.5 \pm 1.8$ \\
		\end{tabular}
	\end{ruledtabular}
\end{table*}

\subsection{Statistical Comparison of BCR and CCR track counts in D$_2$O electrolysis in 0.25 T}

\begin{table*}[!h]
	
	\renewcommand{\arraystretch}{1.2} 
	\setlength{\tabcolsep}{8pt}      

	\caption{\justifying BCR and CCR counts after baseline and background subtraction, derived from the data presented in Table~\ref{tab:s1}. Some CCR entries acquire small negative values and represents statistical residuals arising when the measured track density in a given region falls slightly below the independently determined CCR background value, therefore indicating the absence of a measurable excess above background rather than any unphysical effect. Each data point in this table shares a common standard error of $0.7$. These values are used to perform a two-sample Welch $t$ test.}
	\label{tab:s2}
	\begin{ruledtabular}
		\begin{tabular}{ccccccc}
			Trial & Detector & LT & RT & RB & LB \\
			\hline
			1 & BCR & 369.0 & 418.0 & 440.0 & 426.0 \\
			& CCR & 5.2 & 3.2 & 6.2 & 8.2 \\
			2 & BCR & 366.0 & 331.0 & 340.0 & 416.0 \\
			& CCR & $-0.9$ & 6.2 & 5.2 & 0.2 \\
			3 & BCR & 368.0 & 402.0 & 326.0 & 386.0  \\
			& CCR & 2.2 & 2.2 & 3.2 & 3.2  \\
			4 & BCR & 328.0 & 366.0 & 347.0 & 322.0  \\
			& CCR & 3.2 & 3.2 & $-1.9$ & 4.2 \\
			5 & BCR & 323.0 & 394.0 & 379.0 & 383.0  \\
			& CCR & 0.2 & 2.2 & 9.2 & 7.2  \\
		\end{tabular}
	\end{ruledtabular}
\end{table*}

\twocolumngrid
For the Welch two-sample $t$-test, the null hypothesis assumes that the mean responses of the BCR and CCR detectors are equal. The sample means and variances obtained from the 20 independent BCR and 20 CCR measurements listed in Table~\ref{tab:s2} are:

\begin{equation}
	\begin{aligned}
		\bar{x}_{BCR} &= 371.50, \quad \bar{x}_{CCR} = 3.59, \\
		s_{BCR}^2 &= 1285.85, \quad s_{CCR}^2 = 8.23
	\end{aligned}
\end{equation}

where $\bar{x}_{BCR}$ \& $\bar{x}_{CCR}$ denote the sample mean of BCR and CCR values, respectively, and $s_{BCR}^2$ \& $s_{CCR}^2$ are the corresponding variances. The test statistic is given by $t_{test\ sample}$ and $\nu$, the effective degrees of freedom (Welch--Satterthwaite approximation) gives:

\begin{equation}
	t_{test\ sample} = \frac{\bar{x}_{BCR} - \bar{x}_{CCR}}{\sqrt{\frac{s_{BCR}^2}{n_{BCR}} + \frac{s_{CCR}^2}{n_{CCR}}}} = 45.74
	\label{eq:t_stat}
\end{equation}

\begin{equation}
	\nu = \frac{\left(\frac{s_{BCR}^2}{n_{BCR}} + \frac{s_{CCR}^2}{n_{CCR}}\right)^2}{\frac{\left(\frac{s_{BCR}^2}{n_{BCR}}\right)^2}{n_{BCR}-1} + \frac{\left(\frac{s_{CCR}^2}{n_{CCR}}\right)^2}{n_{CCR}-1}} = 19.24
	\label{eq:nu}
\end{equation}

\textbf{Statistical significance:} For $t_{test\ sample} = 45.7$ with $\nu \approx 19.2 $, the two-tailed test with significance level ($\alpha=0.001$) gives a critical $t$ value ($t_{critical}$) of 3.883, which is far less than the $t_{test\ sample} = 45.7$, ($t_{test\ sample} / t_{critical} \approx 11.7$, and so $p \ll 0.001$). Hence the hypothesis that the BCR and CCR dataset having same mean tracks is rejected. The distribution of BCR and CCR tracks are significantly different.

\onecolumngrid
\subsection{BCR and CCR Track Distributions Recorded during D$_2$O Electrolysis in absence of Magnetic Field and Their Statistical Comparison}

\begin{center}
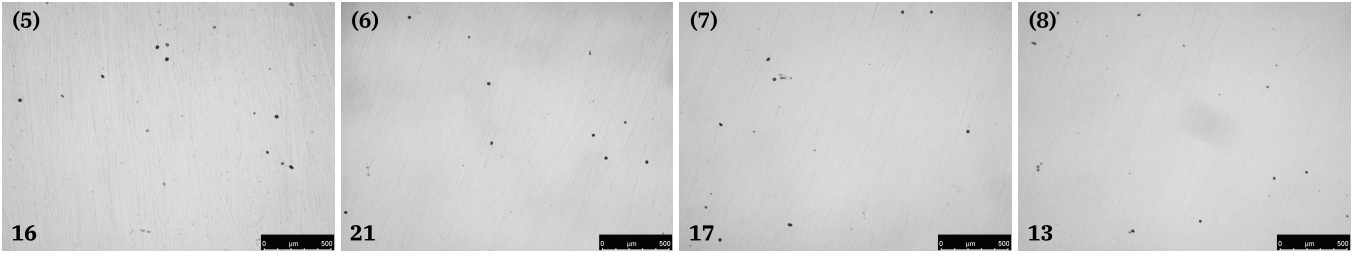

	\includegraphics[width=\textwidth]{CR39/s6.1.pdf}\\[0.2em]
	\includegraphics[width=\textwidth]{CR39/s6.2.pdf}

	\captionsetup{type=figure, width=\linewidth, justification=justified, format=plain}
	\captionof{figure}{\justifying 
	Optical micrographs of a BCR and CCR exposed to D$_2$O electrolysis in absence of magnetic field. Each row of image, from left to right represents LT, RT, RB, and LB region of a BCR or CCR sample. Image labels correspond to: BCR (1–4), CCR (5–8). Track counts of each image is indicated at its bottom left corner.}
	\label{fig:s7}
\end{center}

\subsection{Statistical Comparison of BCR and CCR track counts in D$_2$O electrolysis in absence of any external magnetic field}

\begin{table*}[!h]
	
	\renewcommand{\arraystretch}{1.2} 
	\setlength{\tabcolsep}{8pt}      
	
	\caption{\justifying BCR and CCR track counts after baseline and background subtraction from the raw data presented in Fig.~\ref{fig:s7}. Each data point in this table shares a common standard error of $0.7$. These values are used for the Welch two-sample $t$-test analysis of the BCR and CCR data sets.}
	\label{tab:s3}
	\begin{ruledtabular}
		\begin{tabular}{lccccc}
			& LT & RT & RB & LB \\
			\hline
			BCR tracks & $84.0$ & $82.0$ & $73.0$ & $47.0$ \\
			CCR tracks & $3.2$ & $8.2$ & $4.2$ & $0.2$ \\
		\end{tabular}
	\end{ruledtabular}
\end{table*}

\twocolumngrid
For the Welch two-sample $t$-test, the null hypothesis assumes that the mean responses of the BCR and CCR detectors are equal. The sample means and variances of BCR and CCR values as obtained from Table~\ref{tab:s3}, are:

\begin{equation}
	\begin{aligned}
		\bar{x}_{BCR} &= 71.50, \quad \bar{x}_{CCR} = 3.95, \\
		s_{BCR}^2 &=289.67, \quad s_{CCR}^2 = 10.70
	\end{aligned}
\end{equation}




Using the definitions for the test statistic and effective degrees of freedom given in Eqs.~(\ref{eq:t_stat}) and (\ref{eq:nu}), we calculate:

\begin{equation}
	t_{test\ sample} \approx 7.80 \quad \text{and} \quad \nu \approx 3.29
\end{equation}

\textbf{Statistical significance:} For $t_{test\ sample} = 7.80$ with an effective degree of freedom $\nu \approx 3.2 $, the test statistic exceeds the two-tailed critical value at $\alpha=0.01$, but remains below that at $\alpha=0.002$. This implies that the $p$ value lies in the range $0.002 < p < 0.01$. Hence, the hypothesis that the BCR and CCR dataset having same mean tracks is rejected, and it is confirmed that the BCR and CCR track distributions are significantly different.

\onecolumngrid
\subsection{Bare CR-39 Detector Response to Laboratory Background}

\begin{center}
	\includegraphics[width=\textwidth]{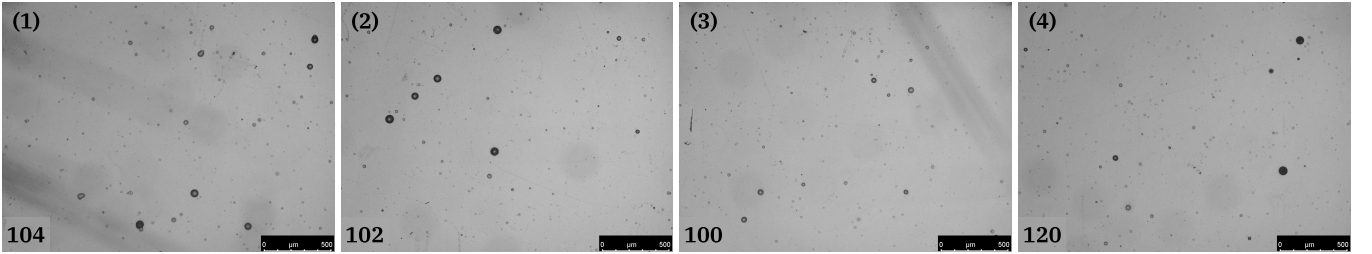}
	
	\captionsetup{type=figure, width=\linewidth, justification=justified, format=plain}
	\captionof{figure}{\justifying Optical micrograph of a bare CR-39 detector exposed to laboratory ambient radiation for one week recorded track counts of (1). 104 (LT), (2). 102 (RT), (3). 100 (RB) and (4). 120 (LB), yielding a mean value of $106.5 \pm 4.0$ tracks. After baseline subtraction, this corresponds to $94.9 \pm 4.1$ tracks/week or equivalently ($0.047 \pm 0.002$) tracks/5 min.}
	\label{fig:s8}
\end{center}
\twocolumngrid

A bare CR-39 sample was placed on the experimental worktable, where detector preparation, electrolysis, etching, and imaging were routinely performed, and left exposed to the laboratory environment for a week.

Figure~\ref{fig:s8} data indicates that a bare CR-39 detector records ($0.047 \pm 0.002$) tracks during a 5 min exposure, corresponding to the maximum time for which the BCR and CCR detectors are handled in bare form and exposed to laboratory background radiation [as explained in Sec.~I~E of the Supplemental Material]. Moreover, it is important to note that any background accumulation during routine handling is a common event to both CR-39 based experimental and background measurements and therefore cancels upon background subtraction.

\onecolumngrid
\subsection{BCR and CCR Detector Calibration Data}

\begin{center}
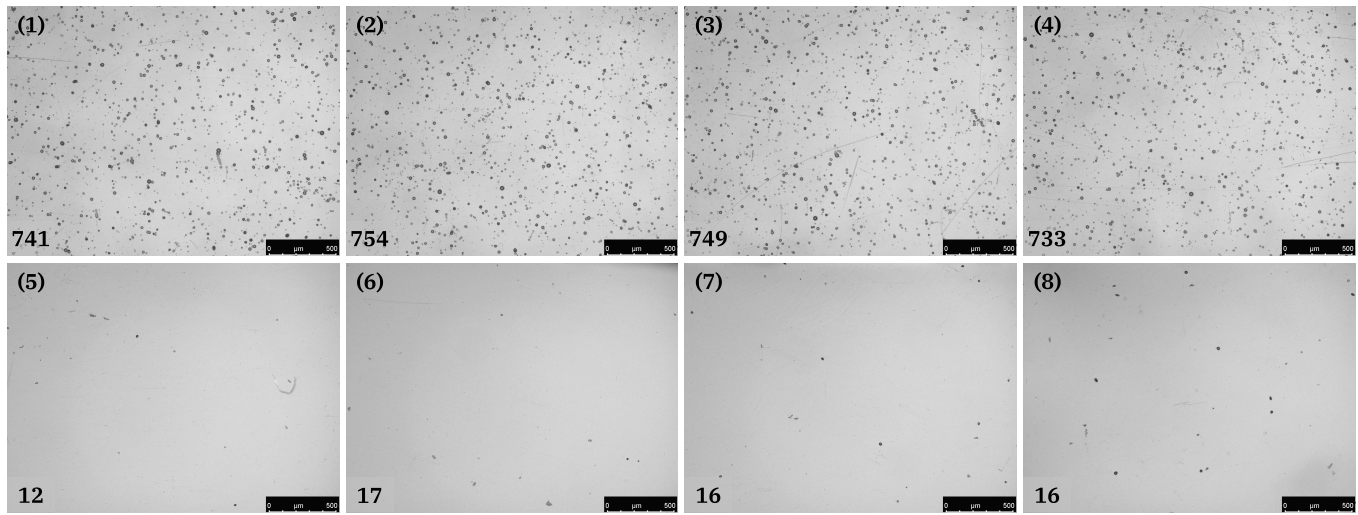

	\includegraphics[width=\textwidth]{CR39/s8.1.pdf}\\[0.2em]
	\includegraphics[width=\textwidth]{CR39/s8.2.pdf}
	
	\captionsetup{type=figure, width=\linewidth, justification=justified, format=plain}
	\captionof{figure}{\justifying Optical micrographs of tracks registered by a BCR  and CCR placed in a thermal neutron reference field. BCR registered track counts of (1). 741 (LT), (2). 754 (RT), (3). 749 (RB) and (4). 733 (LB), yielding a mean value of $744 \pm 4$ tracks. After subtraction of the baseline contribution, this corresponds to $729\pm 4$ tracks accumulated during ($120 \pm 1$) minutes of thermal-neutron exposure. In contrast, the CCR detector exhibited track counts of (5). 12 (LT), (6). 17 (RT) (7). 16 (RB) and (8). 16 (LB), with a mean of $15\pm 1$, which after baseline and background deductions becomes negligible. We note that during $7200 \pm 60$ secs, the total thermal neutrons passing per $4.8$~mm$^2$ area of CR-39 is $\sim 381600 \pm 3180$.}
	\label{fig:s9}
\end{center}

\twocolumngrid

BCR and CCR detectors were imaged after exposure to a common thermal neutron reference field with flux of $1.27 \times 10^{-3}$~cm$^{-2}$s$^{-1}$ for ($120 \pm 1$) minutes [See Fig.~\ref{fig:s9} above].

The detection efficiency of BCR is calculated by taking the ratio of the particle tracks recorded on BCR during exposure ($729 \pm 4$), with the total number of thermal neutrons passing through it in $7200 \pm 60$ secs ($381600 \pm 3180$ neutrons), provides BCR thermal-neutron detection efficiency, $\varepsilon$ $\sim (1.91 \pm 0.02) \times 10^{-3}$. This calibrated efficiency is used to convert measured track densities into absolute slow-neutron fluences in this study.

We emphasize that preparation-related thickness non-uniformity in the boron layer of the BCR can only reduce its effective neutron detection efficiency, thereby suppressing any excess of BCR tracks over CCR observed during neutron detection, and cannot artificially enhance the response. Locally thicker regions will attenuate the charged products of the $^{10}$B(n,$\alpha$)$^{7}$Li reaction before they reach the CR-39, while thinner regions could be insufficiently sensitive to neutrons. Consequently, the measured BCR signal always represents a conservative estimate of the neutron-induced response.

\subsection{Order-of-magnitude estimate of the activity of a hypothetical wired neutron source placed at the center of the cell generating an equivalent neutron flux at BCR in D$_2$O electrolysis (Subject to approximations discussed below)}

The five D$_2$O experimental runs in presence of magnetic field, provides a mean BCR track count of, $N_{\mathrm{trk}}$ = $372 \pm 11$ over an area, $A$ = $0.048$~cm$^2$ during a time period of, $T$ = $604800$ sec (1 week), see Fig.~\ref{fig:3} in main text. Thermal neutron flux, $\Phi$ = $(6.7 \pm 0.2)\, \mathrm{cm^{-2}\,s^{-1}}$, was observed on BCR at the detection site (inner curved surface area of the 25 mL beaker) using $\Phi = N_{\mathrm{trk}} / (\varepsilon\,A\,T)$. Let $S$ be the activity of the hypothetical active wire placed at the center of cell, and $\Phi$ be the flux due to it at any point, then:

\begin{equation}
	S = \oiint \Phi \, dA
\end{equation}

Under the assumption of an approximately constant flux $\Phi$ over the electrolyte’s outer interface (cylindrical shape), the total activity may be estimated as $S \approx \Phi A$, where $A$ is the total surface area of electrolyte’s outer interface [cylinder with radius ($r$) = 1.5 cm and height ($h$) 2.5 cm],

\begin{equation}
	A = [2(\pi)rh + 2(\pi)rl] = 37.70~\text{cm}^2
\end{equation}

Hence, $S = (253 \pm 8)\,\mathrm{neutrons\,s^{-1}}$, corresponding to an equivalent activity of $(6.8 \pm 0.2)$~nano-Curie.

\onecolumngrid
\subsection{BCR and CCR Track Distributions Recorded during H$_2$O Electrolysis under an Applied 0.25 T Magnetic Field}

\begin{center}
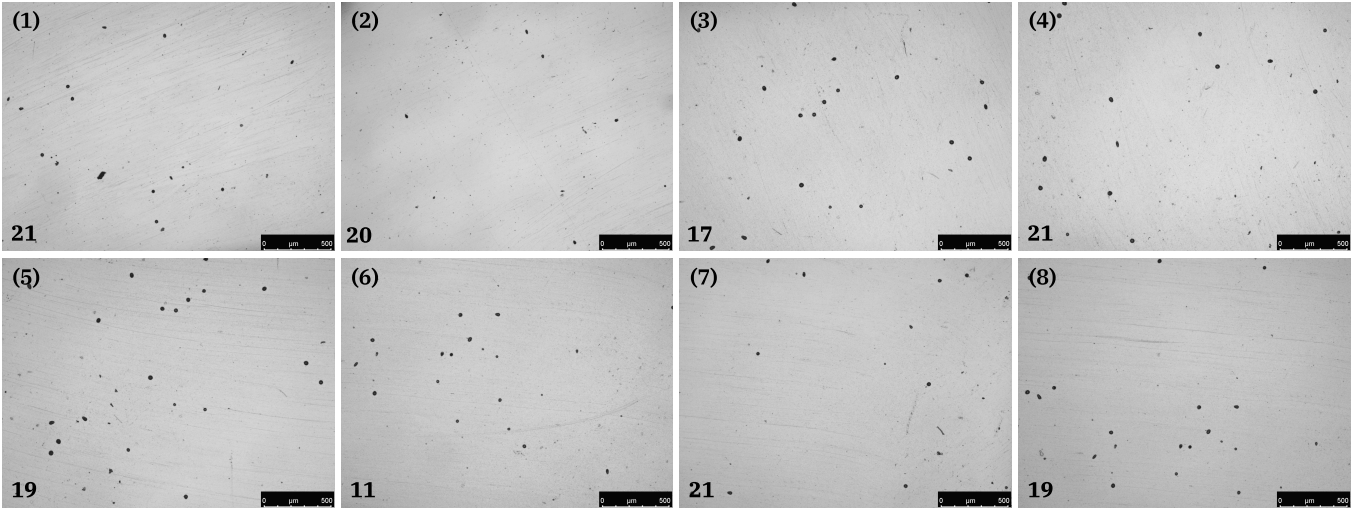

	\includegraphics[width=\textwidth]{CR39/s9.1.pdf}\\[0.2em]
	\includegraphics[width=\textwidth]{CR39/s9.2.pdf}
	
	\captionsetup{type=figure, width=\linewidth, justification=justified, format=plain}
	\captionof{figure}{\justifying Optical micrographs of tracks registered by a BCR  and CCR during H$_2$O electrolysis in presence of 0.25 T. Image label corresponds to BCR: 1.(LT), 2.(RT), 3.(RB) and 4.(LB) and CCR: 5.(LT), 6.(RT), 7.(RB) and 8.(LB).}
	\label{fig:s10}
\end{center}

\subsection{Effect of plastic shielding on bare CR-39 and insensitivity of CR-39 to $\beta/\gamma$ radiation}

\begin{center}
	\includegraphics[width=0.25\textwidth]{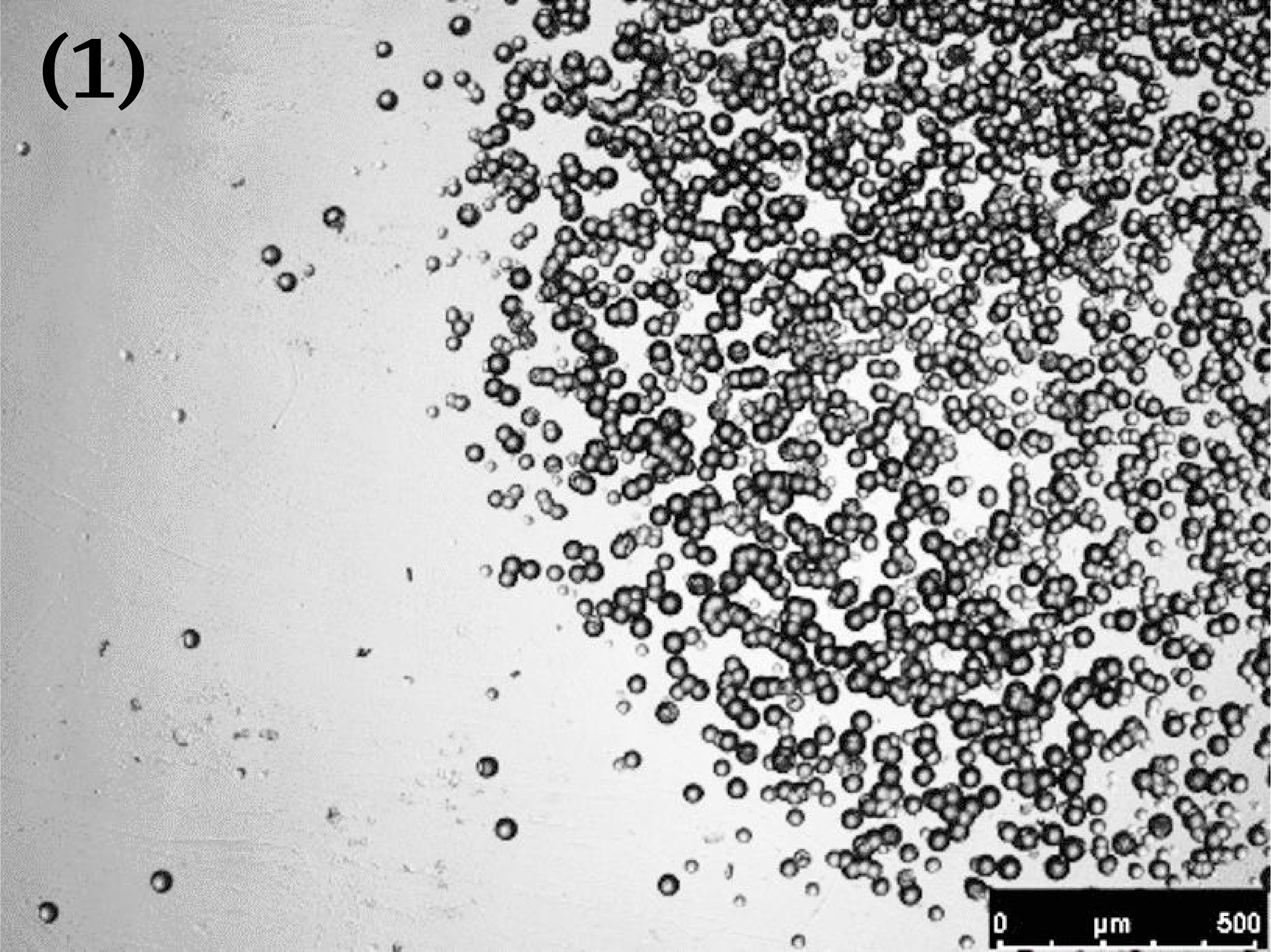}
	\includegraphics[width=0.25\textwidth]{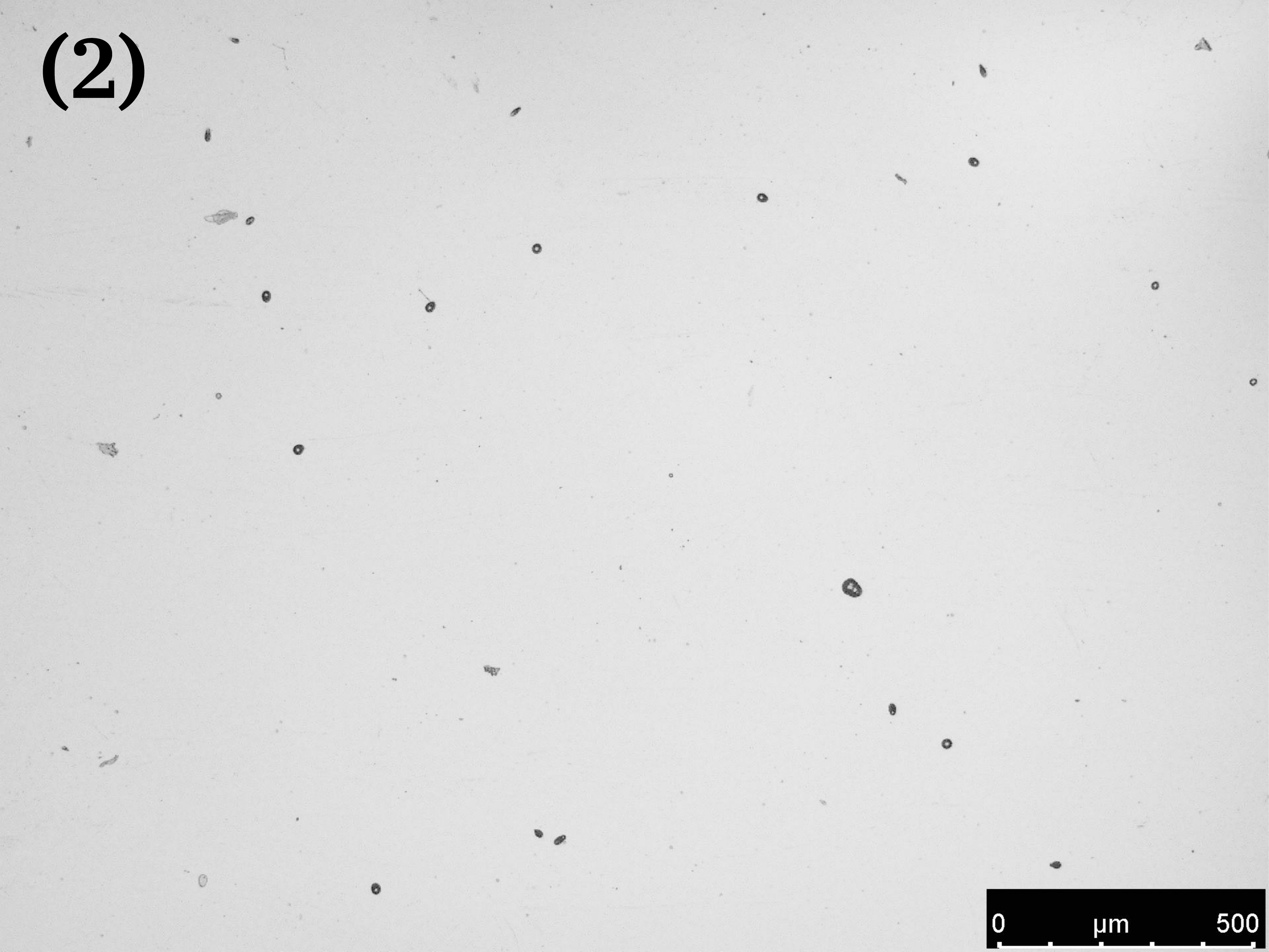}
	\includegraphics[width=0.25\textwidth]{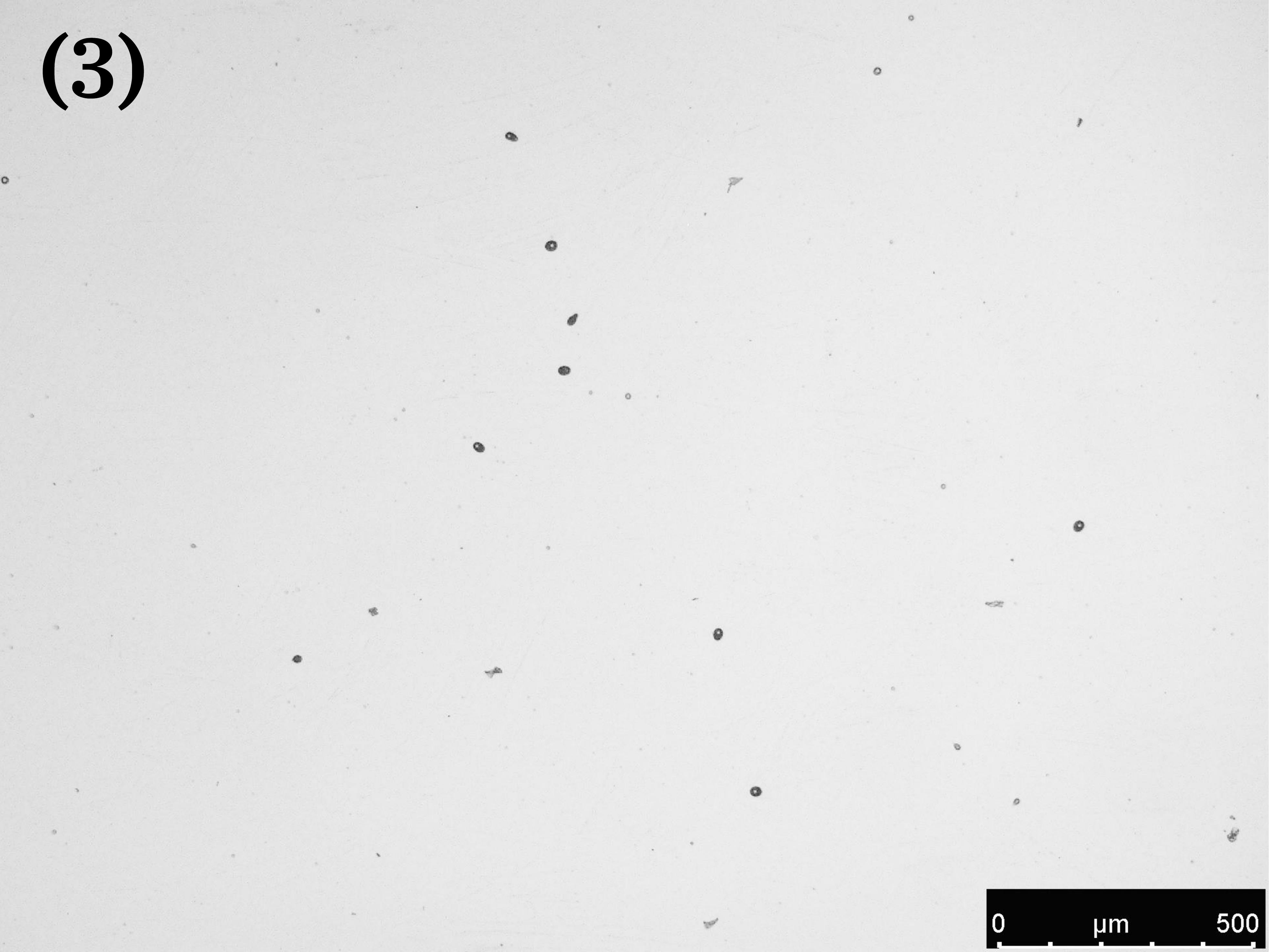}
	
	\captionsetup{type=figure, width=\linewidth, justification=justified, format=plain}
	\captionof{figure}{\justifying optical micrographs of CR-39 based detector exposed to charged-particle, beta, and gamma sources.
		(1) A bare CR-39 detector placed in contact with a 5.48~MeV  $^{241}$Am nCi $\alpha$-particle source for few seconds records a high density of well-defined charged-particle tracks.
		(2) Under identical exposure conditions, no discernible tracks are observed when the CR-39 is covered with a $42~\mu$m polypropylene absorber, demonstrating complete suppression of $\alpha$-particle tracks, consistent with the particle range in polypropylene calculated using SRIM ~\cite{ziegler2010srim} .(3) Representative optical micrograph of a bare CR-39 detector exposed for one hour each to a $^{90}$Sr $\beta$ source and to $^{137}$Cs and $^{60}$Co $\gamma$ sources, showing no discernible tracks above the baseline level, consistent with the detector’s insensitivity to $\beta/\gamma$ radiation under these conditions.}
	\label{fig:s11}
\end{center}



\end{document}